\documentclass{aa}  
\bibpunct{(}{)}{;}{a}{}{,}

\usepackage{subcaption}
\usepackage{afterpage,natbib,lipsum,ulem}
\usepackage{graphicx}
\usepackage[usenames, dvipsnames]{color}
\usepackage{subcaption}
\newcommand{\lya}{Ly-$\alpha$ }

\newcommand{\mvec}[1]{\mathbf{#1}}

\usepackage{txfonts} 
\begin{document}

   \title{Inferring high-redshift large-scale structure dynamics from the Lyman-$\alpha$ forest}


   \author{Natalia Porqueres\inst{1,2} \and Jens Jasche\inst{3} \and Guilhem Lavaux\inst{4,5} \and Torsten En{\ss}lin\inst{1,6}} 

   \institute{ Max-Planck-Institute f\"{u}r Astrophysik (MPA),
             Karl-Schwarzschild-Strasse 1, D-85741 Garching, Germany
         \and
        Excellence Cluster Universe, Technische Universität München, Boltzmannstrasse 2, 85748 Garching, Germany 
         \and
         The Oskar Klein Centre, Department of Physics, Stockholm University, Albanova University Center, SE 106 91 Stockholm, Sweden
         \and
         Institut Lagrange de Paris (ILP), Sorbonne Universit\'{e}s,
98bis boulevard Arago, F-75014 Paris, France
         \and
         Institut d’Astrophysique de Paris (IAP), UMR 7095, CNRS – UPMC Universit\'{e} Paris 6, Sorbonne Universit\'{e}s,
98bis boulevard Arago, F-75014 Paris, France
         \and
         Ludwig-Maximilians-Universität München, Geschwister-Scholl-Platz 1, 80539, München, Germany 
         }

   \date{Received 05/07/2019; accepted 13/09/2019}

 
  \abstract
   {One of the major science goals over the coming decade is to test fundamental physics with probes of the cosmic large-scale structure out to high redshift. Here we present a fully Bayesian approach to infer the three-dimensional cosmic matter distribution and its dynamics at $z>2$ from observations of the Lyman-$\alpha$ forest. We demonstrate that the method recovers the unbiased mass distribution and the correct matter power spectrum at all scales.  Our method infers the three-dimensional density field from a set of one-dimensional spectra, interpolating the information between the lines of sight. We show that our algorithm provides unbiased mass profiles of clusters, becoming an alternative for estimating cluster masses complementary to weak lensing or X-ray observations. The algorithm employs a Hamiltonian Monte Carlo method to generate realizations of initial and evolved density fields and the three-dimensional large-scale flow, revealing the cosmic dynamics at high redshift. The method correctly handles multi-modal parameter distributions, which allow constraining the physics of the intergalactic medium (IGM) with high accuracy. We performed several tests using realistic simulated quasar spectra to test and validate our method. Our results show that detailed and physically plausible inference of three-dimensional large-scale structures at high redshift has become feasible.}
   

   \titlerunning{Inferring high-redshift large-scale structure dynamics from the Lyman alpha forest}
   \authorrunning{Porqueres et al.}
    
   \maketitle
  
%

\section{Introduction}
Currently, cosmology is at a crossroad. While the standard model of cosmology, the $\Lambda$ Cold Dark Matter ($\Lambda$CDM), fits the bulk of cosmological observations to extraordinary accuracy, some tensions between the model and observations seem to persist and increase \citep{Planck15,Planck2018nonGaussianity}. Amongst them are the $H_0$ and $\sigma_8$ tensions \citep[e.g.][]{2016A&A...594A..24P, 2016ApJ...826...56R, 2017MNRAS.471.4412K, 2018ApJ...861..126R, 2018PhRvD..98d3526A, 2019arXiv190509338R}.  Additionally, Lyman-$\alpha$ auto-correlation and cross-correlations with quasar data reported a $2.3\sigma$ tension with the flat $\Lambda$CDM prediction obtained from Planck observations \citep[see e.g.][]{2015A&A...574A..59D,2017A&A...608A.130D}.

The resolution of these tensions may encompass systematic effects, but may also be the first signs of new physics indicated by novel cosmological data of increasing quality. New observations and better control on systematic effects in data analyses are inevitable to gain new insights into the physical processes driving the evolution of the universe. For this reason, in this work, we present a novel and statistically rigorous approach to extract cosmologically relevant and significant information from high-redshift Lyman-$\alpha$ forest observations tracing the dynamic evolution of cosmic structures.

Detailed analyses of the spatial distribution of matter and growth of cosmic structures can provide significant information to discriminate between models of homogeneous dark energy and modifications of gravity, to determine neutrino masses and their mass hierarchy, and to investigate the dynamical clustering behaviour of warm or cold dark matter \citep[e.g.][]{1996ApJ...458....1C, 2015APh....63...23H, 2016PhRvD..94l3525B, 2012AnP...524..507F, 2018JCAP...03..035B, 2018PhRvD..97l3544M}.

To harvest this information, cosmology now turns to analyse the inhomogeneous matter distribution with next-generation galaxy surveys such as the Large Synoptic Survey Telescope (LSST) and the Euclid satellite mission probing the galaxy distribution out to redshifts $z\sim3$ and beyond \citep[][]{2009arXiv0912.0201L, 2010arXiv1001.0061R, 2011arXiv1110.3193L, 2016MNRAS.456.3743A}.

While most of this research focuses on studying the cosmic large-scale structure with galaxy clustering and weak lensing observations, the Lyman-$\alpha$ (Ly\nobreakdash-$\alpha$) forest has the potential to provide important complementary information. Firstly, the Ly\nobreakdash-$\alpha$ forest probes the matter distribution at higher redshift with higher spatial resolution than can be achieved with galaxy sampling rates. By probing scales down to $1$ Mpc, the Ly-$\alpha$ forest is sensitive to neutrino masses and dark matter models \citep[][]{WDMViel, LyaNeutrinos14,NeutrinoLya15,NeutrinoLya15b,NeutrinoLya17,LyaNeutrinos17}. Secondly, while galaxy clustering probes high-density regions, the Ly-$\alpha$ forest is particularly sensitive to under-densities in the matter distribution \citep[][]{Peirani14,Sorini16}. In addition, the Ly-$\alpha$ at $z>2$ is redshifted to an optical band for which the atmosphere is transparent. The Ly-$\alpha$ emission becomes then one of the more powerful probes at redshifts that are otherwise challenging to access with ground-based galaxy surveys \citep{KGLee14}.

Specifically, auto-correlations of the Ly-$\alpha$ forest along the line of sight have been used to infer cosmological parameters   which agree by and large with CMB observations \citep[e.g.][]{CosmoLyaSeljak,CosmoLyaViel,CosmoParamsSlosar,CosmoParamsBusca,CosmoParamsLyaBautista,BAOeBOSSLya}. Ly-$\alpha$ forest data have also been used to constrain neutrino masses  \citep{NeutrinoLya15,NeutrinoLya15b,LyaNeutrinos17}, test warm dark matter models \citep{WDMViel}, and study the thermal history of the intergalactic medium \citep{ReionLyaNasir,ReionLyaBoera}.

In line with these promises, a large number of surveys mapping the Ly-$\alpha$ forest at higher redshifts has been launched. The eBOSS observations in the SDSS DR14 are expected to provide the spectra of 435~000 quasars over 7500~deg$^2$ and re\nobreakdash-observe the lines of sight from the previous data release with low signal-to-noise  \citep[S/N<3,][]{eBOSSMyers}.  DESI \citep[Dark Energy Spectroscopic Instrument,][]{DESI, DESI2} will map the Ly-$\alpha$ forest over a larger fraction of the sky (14~000~deg$^2$), targeting 50 high-redshift quasars per deg$^2$. Increasing the density of tracers, the ongoing COSMOS Lyman Alpha Mapping and Tomography survey \citep[CLAMATO,][]{ClamatoDR1} maps the Ly-$\alpha$ forest over a small part of the sky (600 arcmin$^2$) with a separation between lines of sight of 2.4 h$^{-1}$ Mpc, 20 times smaller than the BOSS survey \citep{SDSS,Dawson13}  and smaller than the expected separation in DESI \citep[3.7h$^{-1}$ Mpc,][]{Horowitz19}. Future surveys like Maunakea Spectrographic Explorer \citep[MSE][]{MSE} and LSST \citep{LSST} will increase the amount of available Ly-$\alpha$ forest observations by a factor of 10. 

\begin{figure}
    \centering
        {\includegraphics[width=\hsize,clip=true]{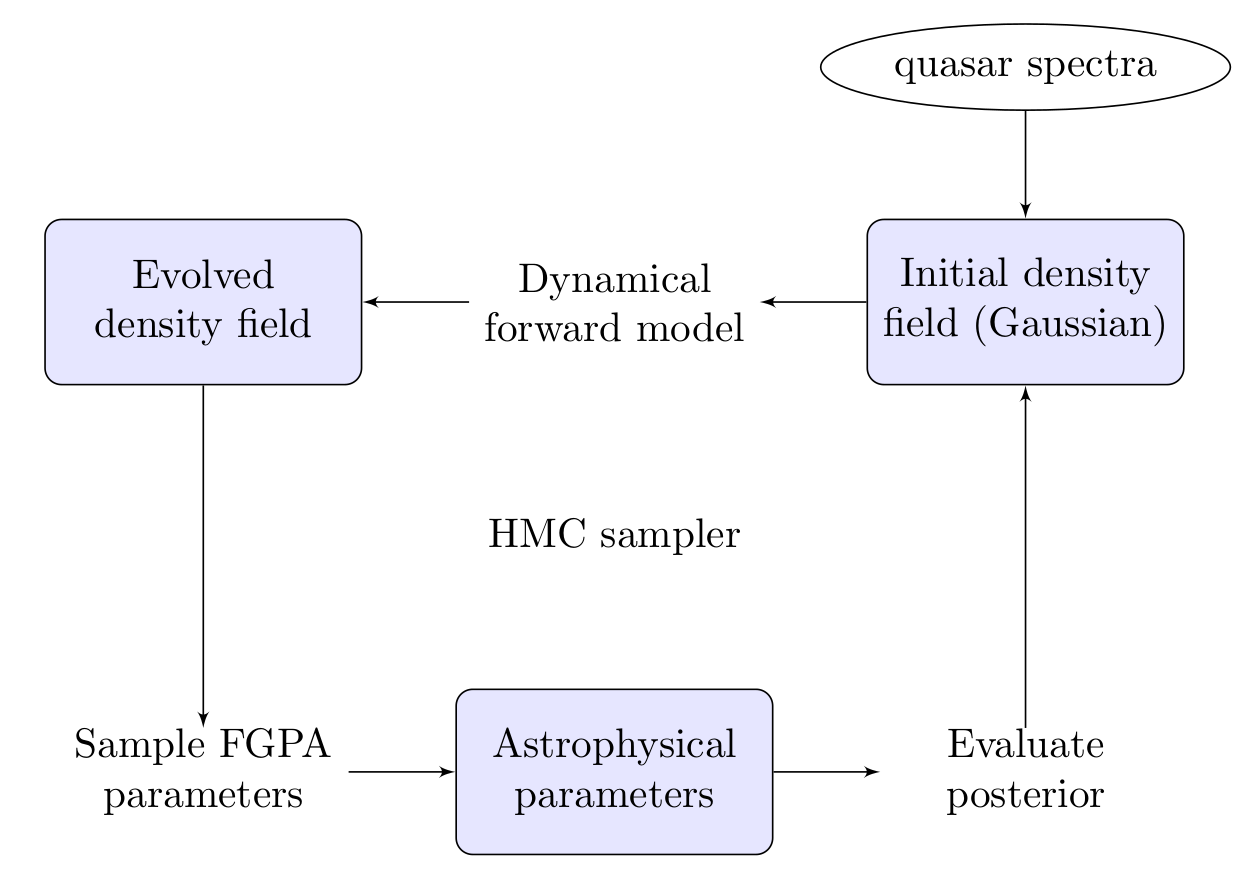}}
    \caption{Flow chart depicting the iterative block sampling approach of the BORG inference framework. As can be seen, the BORG algorithm first infers three-dimensional initial and evolved density fields from quasar spectra using assumed astrophysical parameters of the FGPA model. Then, using realizations of the evolved density field, the algorithm updates these astrophysical parameters and a new iteration of the process begins. Iterating this procedure results in a valid Markov Monte Carlo Chain that correctly explores the joint posterior distribution of the three-dimensional matter distribution and astrophysical properties underlying \lya forest observations.}    
    \label{fig:chart}
\end{figure} 

At present, major analyses of the Ly-$\alpha$ forest focus only on the analysis of the matter power spectrum \citep[e.g.][]{Croft98,CosmoLyaSeljak,CosmoLyaViel,Bird11,NeutrinoLya15,NeutrinoLya15b, ReionLyaNasir, LyaNeutrinos17,ReionLyaBoera}. However, these approaches ignore significant amounts of information contained in the higher-order statistics of the matter density field as generated by non-linear gravitational dynamics in the late time universe \citep[][]{cmb_lensing_filaments}. 

Capturing the full information content of the cosmic large-scale structure requires a field-based approach to infer the entire three-dimensional cosmic large-scale structure from observations.
This poses a particular challenge for the analyses of Ly-$\alpha$ forest observations, which provide sparse inherently one-dimensional information along the lines of sight. Various approaches to perform three-dimensional density reconstructions from one-dimensional Ly-$\alpha$ forests have been proposed in the literature \citep[e.g.][]{Kitaura12,Cisewski14,Stark15,Ozbek16,Horowitz19}. \citet{Gallerani11} and \citet{Kitaura12} proposed a Gibbs sampling scheme to jointly infer density and velocity fields and corresponding power-spectra. However, these approaches assume matter density amplitudes to be log-normally distributed. The log-normal distribution reproduces one- and two-point statistics but fails to reproduce higher-order statistics associated with the filamentary dark matter distribution. In an attempt to extrapolate information from one-dimensional quasar spectra into the three-dimensional volume, \cite{Cisewski14} applied a local polynomial smoothing method. \cite{Ozbek16} and \cite{Stark15} employed a Wiener filtering approach to reconstruct the three-dimensional density field between lines of sight of Ly-$\alpha$ forest data. In order to reproduce higher-order statistics, \cite{Horowitz19} recently used a large-scale optimization approach to fit a gravitational structure growth model to Ly-$\alpha$ data, showing that this approach allows recovering the more filamentary structure of the cosmic web. Although the approach improves over linear and isotropic Wiener filtering approaches, it shows systematic deviations of reconstructed matter power-spectra and underestimates density amplitudes at scales corresponding to the mean separation between lines of sight \citep{Horowitz19}. 

To go beyond previous approaches, in this work, we present a fully Bayesian and statistically rigorous approach to perform dynamical matter clustering analyses with high-redshift Ly-$\alpha$ forest data while accounting for all uncertainties inherent to the observations. The aim of this work is to provide a novel and fully Bayesian approach to infer physically plausible three-dimensional density and velocity fields from Ly-$\alpha$ forest data. Our approach builds upon the algorithm for Bayesian Origin Reconstruction from Galaxies \citep[BORG,][]{BORG,PM}, which employs physical models of structure formation and sophisticated Markov Chain Monte Carlo techniques to optimally extract large-scale structure information from data and quantify corresponding uncertainties. 

To make such inferences feasible, we developed a likelihood based on the fluctuating Gunn-Peterson approximation \citep[FGPA,][]{GPeffect} and jointly constrained the astrophysical properties of the intergalactic medium. We tested and validated our approach with simulated data emulating the CLAMATO survey, showing that the algorithm recovers the unbiased dark-matter field. 

 The paper is organized as follows. Section \ref{BORG} provides a brief overview of our Bayesian inference framework, BORG, as required for this work. In Section \ref{sect:lya_physics}, we discuss the physics of the \lya forest underlying the data model described in Section \ref{sec:data_model_lya}.  Section \ref{sect:mock_data} describes the generation of simulated data employed to test and validate the algorithm. The performance of the algorithm is tested in Section \ref{sec:lya_test} and the results are shown in Section \ref{sect:lya_results}, where we also discuss possible applications of the method to scientifically exploit the Ly-$\alpha$ forest. Finally, Section \ref{conclusions} summarizes the results and discusses further extensions of the algorithm.

\begin{figure}
        \centering
        {\includegraphics[width=\hsize,clip=true]{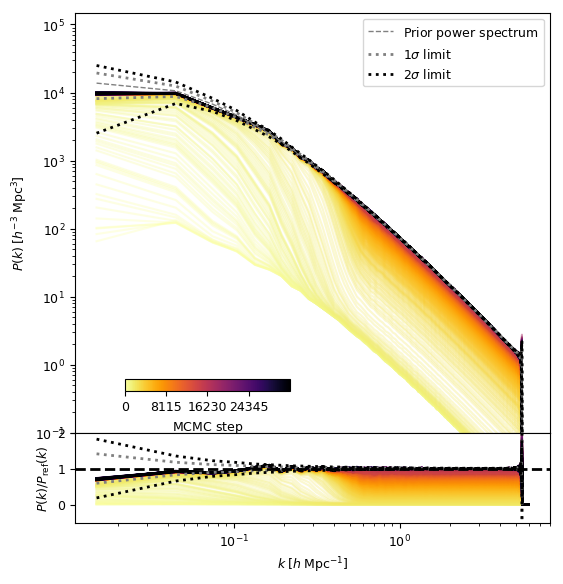}}
        \caption{Burn-in of the posterior  matter power spectra from the inferred initial conditions from a BORG analysis. The colour scale shows the evolution of the matter power spectrum with the number of samples: the Markov chain is initialized with a Gaussian initial density field scaled by a factor $10^{-3}$ and the amplitudes of the power spectrum systematically drift towards the fiducial values. Monitoring this drift allows us to identify when the Markov chain approaches a stationary distribution and provides unbiased estimates of the target distribution. The dashed lines indicate the underlying power spectrum and the 1- and 2-$\sigma$ cosmic variance limits. At the end of the warm-up phase, the algorithm recovers the true matter power spectrum in all range of Fourier modes.}
        \label{fig:pk_burnin}
    \end{figure}

\begin{figure*}
        \centering
            {\includegraphics[width=0.95\hsize,clip=true]{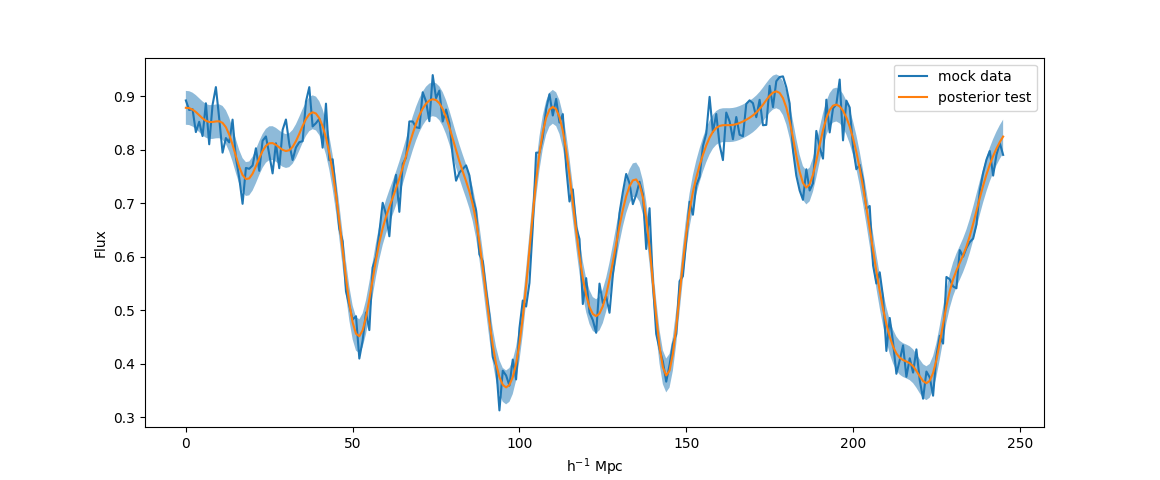}}
         {\includegraphics[width=0.95\hsize,clip=true]{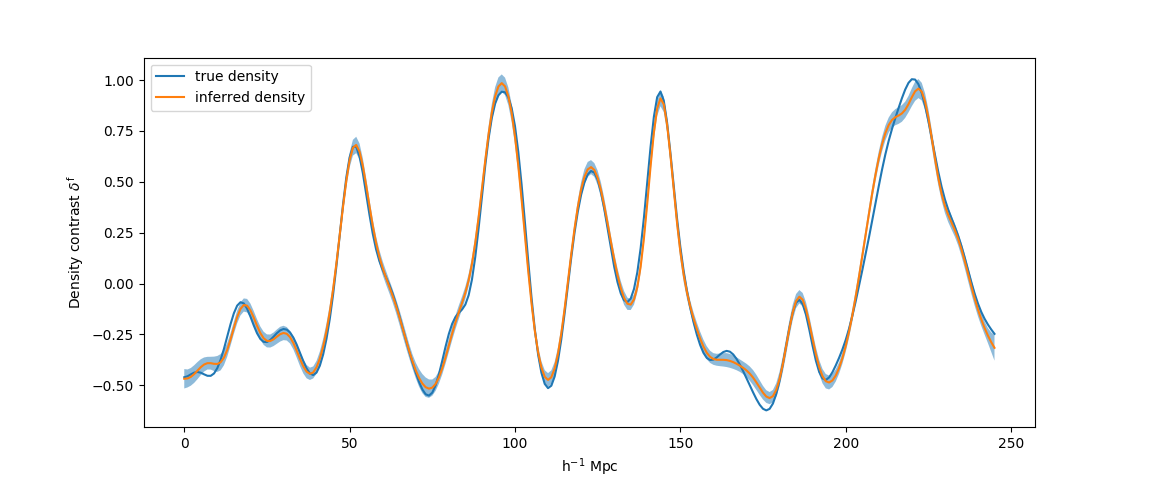}}
        \caption{Top panel: Posterior predictive flux for a spectrum with noise $\sigma = 0.03$. The posterior predicted flux is computed from the ensemble mean density field, shown in the lower panel (orange line). The blue shaded region indicates the 1-$\sigma$ region, corresponding to the standard deviation of the noise in this line of sight. These tests check whether the data model can accurately account for the observations. Any significant mismatch would immediately indicate a breakdown of the applicability of the data model or error of the inference framework. Our method recovers the transmitted flux fraction correctly, confirming that the data model can accurately account for the observations. Bottom panel: Comparison of the inferred ensemble mean density field along the line of sight to the ground truth. It can be seen that high-density regions yield a suppression of transmitted flux, while under-dense regions transmit the quasar signal, in agreement with the FGPA model. The density field has been smoothed with a Gaussian smoothing kernel of $\sigma=0.5 h^{-1}$~Mpc to simulate the difference between dark matter and gas density fields \citep{Peirani14}. This smoothing kernel is applied to the dark matter density field before generating the mock data.
        \label{fig:posterior_test}}
    \end{figure*}

\section{The physics of the Lyman-$\alpha$ forest} 
\label{sect:lya_physics}
Developing a Bayesian framework to infer the matter distribution from the Ly-$\alpha$ forest requires to incorporate the corresponding light absorption physics into the data model. 

When traversing the universe, photons emitted from quasars are scattered by neutral hydrogen (HI) gas residing inside cosmic large-scale structures. Scattering of quasar light results in observed spectra covered with absorption features referred to as the \lya forest. While high-density regions obscure the light and attenuate quasar fluxes, under-dense regions are almost transparent. Therefore, the signal comes from the under-dense regions. Due to the cosmological redshift, different HI regions absorb photons from different wavelengths in the quasar spectrum \citep[see e.g.][]{MoWhite}. Consequently, every HI absorber leaves an absorption feature to the spectrum, permitting to trace the distance and density of HI regions along the line of sight. The \lya forest, therefore, provides a formidable probe of the cosmic large-scale structure along the observers past light cone. 

Since the Ly-$\alpha$ forest generated at $z>2$ is redshifted to the optical atmospheric window, it can be observed by ground-based telescopes, becoming one of the more relevant probes at redshifts that are otherwise challenging to access with galaxy surveys. Observations of the Ly-$\alpha$ forest at lower redshift, from $z=1.5$ down to $z=0$ \citep{Lowz99,Lowz01,LowzLya10}, require UV space-based spectrographs as  Faint Object Spectrograph in the Hubble Space Telescope \citep{HSTLya93,HSTLya96,HSTLya98}.

In contrast to galaxy surveys, requiring assumptions on galaxy formation to model galaxy biases, modelling the Ly-$\alpha$ forest does not involve complicated models. At $z>2$, HI gas, producing the Ly-$\alpha$ forest, is in radiative equilibrium of photoionization due to ultraviolet (UV) background and adiabatic cooling due to the cosmic expansion \citep{LyPhysics}. This yields a tight relation between the temperature of the IGM and the dark matter density $\rho$
\begin{equation}
T = T_0 \Bigg(\frac{\rho}{\bar{\rho}}\Bigg)^{\gamma-1},
\label{eq:T_rho}
\end{equation}
where $T_0$ and $\gamma$ are constants that depend on the re-ionization history and the spectral shape of the UV background sources \citep{LyPhysics}. 

In thermal equilibrium, HI number density can be expressed as \citep{LyPhysics}
\begin{equation}
n_{HI}(\mvec{x}) = \frac{\alpha(T(\mvec{x}))}{\Gamma}\left[n_0(1+z)^3(1+\delta(\mvec{x}))\right]^2 \propto (1+\delta(\mvec{x}))^\beta,
\label{eq:nHI}
\end{equation}
where $\mvec{x}$ is the comoving position, $\alpha(T)\propto T^{-0.7}$ is the radiative recombination rate of HI, $\Gamma$ is the photoionization rate of neutral hydrogen, $n_0(1+z)^3$ is the mean baryon number density at redshift $z$ and $\delta$ is the dark matter density contrast. On scales larger than Jeans' scale ($100$ kpc), thermal pressure in the gas is negligible and the IGM traces the dark matter distribution with high accuracy \citep{Peirani14}. 

Combining equations~\eqref{eq:T_rho} and \eqref{eq:nHI} leads to the fluctuating Gunn-Peterson approximation \citep[FGPA][]{GPeffect} for the transmitted flux:
\begin{equation}
F(z,\mvec{\hat{x}}) = \exp\left[-A(1+\delta(\mvec{x}))^\beta\right],
\label{eq:FGPA}
\end{equation}
with $z$ the considered absorption redshift, $\mvec{x}$ the corresponding comoving distance, $\mvec{\hat{x}}$ the associated unit vector, $\beta = 2-0.7(\gamma-1)$ and $A \propto (1+z)^6 T_0^{-0.7} \Gamma^{-1}$. The connection between the redshift and the cartesian coordinate of the regular grid $\mvec{x}$ is given by the projector along the line of sight (see Appendix \ref{app:projector}.  We note that the astrophysical properties of HI gas are encoded in the parameters $A$ and $\beta$ of the FGPA model. Although this work does not include a treatment of redshift space distortions (RSD) or thermal broadening of the absorption lines, we intend to extend the current formalism to include it in future works.

\section{A data model for the \lya forest}
\label{sec:data_model_lya}
To incorporate Ly-$\alpha$ forest observations into the Bayesian framework, we  now built a data model using the light absorption physics described in Sect. \ref{sect:lya_physics}. Specifically, we assume the FGPA transmission model and Gaussian pixel noise for measured quasar spectra. A corresponding likelihood distribution can then be expressed as:
\begin{multline}
P(\delta^\mathrm{f}|F) = \prod_{n,x} \frac{1}{\sqrt{2\pi}\sigma} \\ 
 \times \exp\left[-\frac{\left((F_n)_x - \exp\left[-A(1+\delta^\mathrm{f}_x)^\beta\right]\right)^2}{2\sigma^2}\right],
\label{eq:likelihood_sect}
\end{multline}
where $n$ labels respective quasar spectra, $x$ indexes different volume elements intersected by the $n$-th quasar line of sight and $\delta^\mathrm{f}$ is the non-linear final density contrast evaluated at $z=2.5$.

This likelihood is then implemented into the large-scale structure sampler of the BORG framework. The corresponding physical forward modelling approach proceeds as follows. Using realizations of the three-dimensional field of primordial fluctuations, the dynamical structure formation model evaluates non-linear realizations of the dark matter distribution at $z=2.5$. Using these matter field realizations and the FGPA data model, BORG predicts quasar spectra that are compared to actual observations via the Gaussian likelihood of equation~\eqref{eq:likelihood_sect}. More details on the sampling procedure are given in Appendix~\ref{method}. To efficiently implement this non-linear and non-Gaussian forward modelling approach into an MCMC framework, we employed a Hamiltonian Monte Carlo (HMC) method, as described in Appendix~\ref{hmc}.

In this work, we focus on the inference of the spatial dark matter distribution and its dynamics. For this reason, we treat the parameters $A$ and $\beta$ mostly as nuisance parameters. Following similar approaches described in previous works \citep{BORG,Foregrounds,Altair,PM}, we use efficient slice sampling techniques to sample these parameters. We note that these parameters can carry valuable information on the astrophysical properties of the IGM. As illustrated in section \ref{sec:meta_results_lya}, our inference can also be used to make statements on the astrophysics of the IGM.

The entire iterative sampling approach is depicted in Figure \ref{fig:chart}. As can be seen, the method iteratively infers three-dimensional matter density fields and parameters of the FGPA model, resulting in a valid MCMC approach to jointly explore density fields and astrophysical parameters.   
\begin{figure}
        \centering
            {\includegraphics[width=\hsize,clip=true]{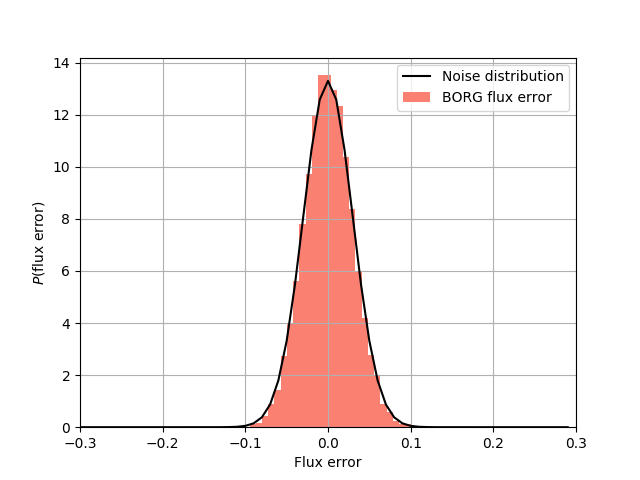}}
        \caption{Histogram of the error in the fractional transmitted flux, which is computed as the difference between posterior-predicted fluxes and input spectra. The distribution of flux error matches the distribution of pixel noise in the data, indicating that the method is close to the theoretical optimum. This distribution of error flux can be compared to the one obtained in previous works, demonstrating that our method recovers the fluxes with significantly higher accuracy.}\label{fig:flux_error}
    \end{figure}

   \begin{figure}[h]
    \centering
            {\includegraphics[width=\hsize,clip=true]{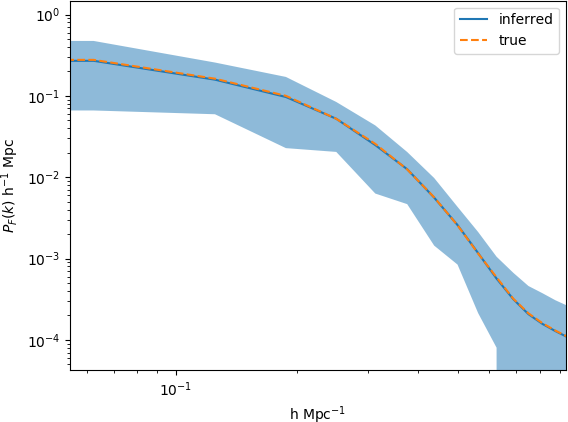}}
        \caption{One-dimensional power spectrum of the flux. The orange line corresponds to the mean flux power spectrum computed from posterior-predicted spectra. The blue-shaded region corresponds to the standard deviation between the different lines of sight. The blue line indicates the flux power spectrum for the mock data. This test shows that the algorithm recovers the flux spectrum inside the $1-\sigma$ uncertainty limit.
        }\label{fig:pk_flux}
    \end{figure}

\section{Method}
\label{BORG}
As mentioned above, this work extends our previously developed BORG algorithm in order to analyse the spatial matter distribution underlying \lya forest observations. Here we provide only a brief summary of the algorithm and corresponding concepts, but interested readers can find more detailed descriptions in our previous works \citep[][]{BORG,JLW15,2MPP,Foregrounds,PM}.

The BORG algorithm is a large-scale Bayesian inference framework aiming at inferring the non-linear spatial dark matter distribution and its dynamics from cosmological data sets. The underlying idea is to fit full dynamical gravitational structure formation models to observations tracing the spatial distribution of matter. Using non-linear structure growth models, the BORG algorithm can exploit the full statistical power of higher-order statistics imprinted to the matter distribution by gravitational clustering. In fitting dynamical structure growth models to data, the task of inferring non-linear matter density fields turns into a statistical initial conditions problem aiming at inferring the spatial distribution of primordial matter fluctuations from which present structures formed via gravitational structure growth. As such the BORG algorithm naturally links primordial initial conditions to late time observations and permits to infer dynamical properties and the structure formation history from present galaxy observations \citep[][]{BORG,JLW15,2MPP,Foregrounds,PM}.  

The BORG algorithm employs a large-scale structure posterior distribution based on the well-developed prior understanding of almost Gaussian primordial density fluctuations and gravitational structure formation to predict physically plausible realizations of present matter density fields. More specifically BORG encodes a Gaussian prior for the initial density contrast at an initial cosmic scale factor of $a = 10^{-3}$. Initial and evolved density fields are linked by deterministic gravitational evolution mediated by various physics models of structure growth. Specifically, BORG incorporates physical models based on Lagrangian Perturbation Theory (LPT) but also fully non-linear particle mesh (PM) models. 

Our Bayesian inference approach has been previously shown to perform accurate dynamic mass estimation and provide mass measurements that agree well with complimentary standard weak lensing and X-ray observations \citep[][]{PM}. More recently, we have shown, that BORG can exploit the geometric shapes of the cosmic large-scale structure to significantly improve constraints on cosmological parameters via the Alcock-Paczynski test \citep[][]{Altair}. Additional projects, conducted with the BORG algorithm, can be found at our web-page of the Aquila consortium\footnote{\url{https://aquila-consortium.org}}.

In this work, we rely on an LPT approach to structure formation since the Ly-$\alpha$ forest mostly arises from under-dense regions that can be conveniently modelled by perturbation theory \citep{Peirani14,Sorini16}.
The dynamical model permits to recover the non-linear higher-order statistics associated with the filamentary matter distribution of the cosmic large-scale structure. Our approach also immediately provides detailed inferences of large-scale velocity fields and structure growth information at high redshifts as is demonstrated in the remainder of this work.

A feature of particular relevance to this work is that BORG employs a modular statistical programming engine that executes a statistically rigorous Markov Chain to marginalize out any nuisance parameters associated to the data model, such as unknown biases or systematic effects. This statistical programming engine permits us to easily and straightforwardly implement complex data models. In particular, in this work, we perform joint analyses of the cosmological large-scale structure and unknown astrophysical properties of the IGM medium described by the parameters of the FGPA model. The corresponding iterative block sampling inference approach is illustrated in Fig. \ref{fig:chart}.

    \begin{figure}[t]
        \centering
            {\includegraphics[width=\hsize,clip=true]{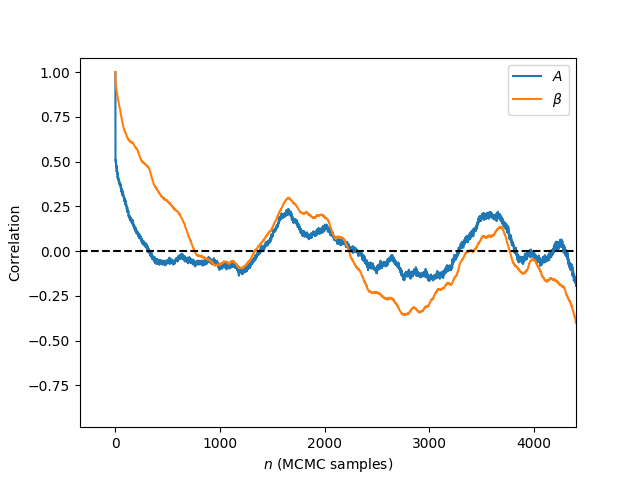}}
        \caption{Autocorrelation of the parameters as a function of the sample lag in the Markov chain. The correlation length of the sampler can be estimated by determining the point when correlations drop below 0.1 for the first time. These parameters have a correlation length of 830 samples for $\beta$ and $640$ for $A$.}\label{fig:meta_corr}
    \end{figure}
    
\section{Generating artificial Ly-$\alpha$ forest observations}
\label{sect:mock_data}
To test our Ly-$\alpha$ forest inference framework, we generate artificial mock observations emulating the CLAMATO survey \citep{ClamatoMock,ClamatoDR1}.

Mock data are constructed by first generating Gaussian initial conditions on a cubic Cartesian grid of side length of 256$h^{-1}$~Mpc with a resolution of 1$h^{-1}$~Mpc. To generate primordial Gaussian density fluctuations we use a cosmological matter power-spectrum including the Baryonic wiggles calculated according to the prescription provided by \cite[][]{EH98,EH99}. We further assume a standard $\Lambda$CDM cosmology with the following set of parameters: $\Omega_m = 0.31,\ \Omega_\Lambda = 0.69,\ \Omega_b = 0.022,\ h=0.6777,\ \sigma_8= 0.83,\ n_s = 0.9611$ \citep{Planck15}. We assumed $H=100 h$~km~s$^{-1}$~Mpc$^{-1}$.

To generate realizations of the non-linear density field, we evolve these Gaussian initial conditions via Lagrangian Perturbation Theory. This involves simulating displacements for $512^3$ particles in the LPT simulation. Final density fields are constructed by estimating densities via the cloud-in-cell scheme from simulated particles on a Cartesian equidistant grid with $256^3$ volume elements. We further apply a Gaussian smoothing kernel of $\sigma=0.5 h^{-1}$~Mpc to the fields to simulate the difference between dark matter and gas density fields \citep[see e.g.][]{Peirani14,ClamatoMock}. A three dimensional quasar flux field is generated by applying the FGPA model, given in equation (\ref{eq:FGPA}), to the final density field and assuming constant parameters $A = 0.35$ and $\beta=1.56$ at $z=2.5$, corresponding to the values in \cite{ClamatoMock}. 

From this three-dimensional quasar flux field, we generate individually observed skewers by tracing lines of sight through the volume. 
Specifically, we generate a total of $1024$ lines of sight parallel to the $z$-axis of the box, distributed in a regular grid of $32 \times 32$ lines of sight spaced regularly with a separation of  8 h$^{-1}$ Mpc.  The separation between lines of sight is the most limiting factor of Ly-$\alpha$ surveys. Although the CLAMATO survey achieves a separation of 2.4 h$^{-1}$ Mpc,  the lines of sight in this work are separated by 8 h$^{-1}$ Mpc. This gives us the opportunity to explore the potential of our method to reconstruct the three-dimensional density field between lines of sight. Finally, we added Gaussian pixel-noise to the flux with $\sigma = 0.03$. 

Although tests are performed with the same noise distribution in all lines of sight, Section \ref{sec:snr} presents an analysis with a different signal-to-noise ratio (S/N) for each line of sight.

         \begin{figure}[t]
        \centering
            {\includegraphics[width=\hsize,clip=true]{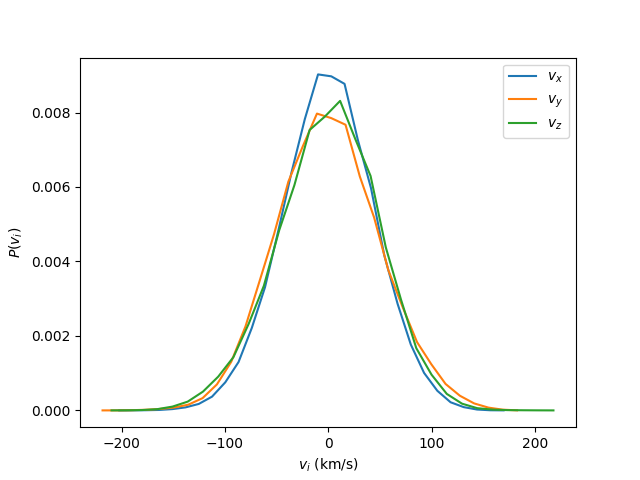}}
        \caption{Distribution of the three Cartesian components of the velocity. The three distributions have the same mean and variance, indicating that the density field is isotropic as expected.}\label{fig:velocities}
    \end{figure}
    
\section{Testing sampler performance}
\label{sec:lya_test}
In this section, we present a series of tests to evaluate the performance of our method. In particular, we focus on the convergence (Section~\ref{sec::warmup}) and efficiency (Section~\ref{sec:efficiency}) to infer the underlying dark matter density field. We provide tests of the posterior power-spectra and perform posterior predictive tests for quasar spectra to check that the inferred model agrees with the data.

\begin{figure*}
    \centering
        {\includegraphics[width=\hsize,clip=true]{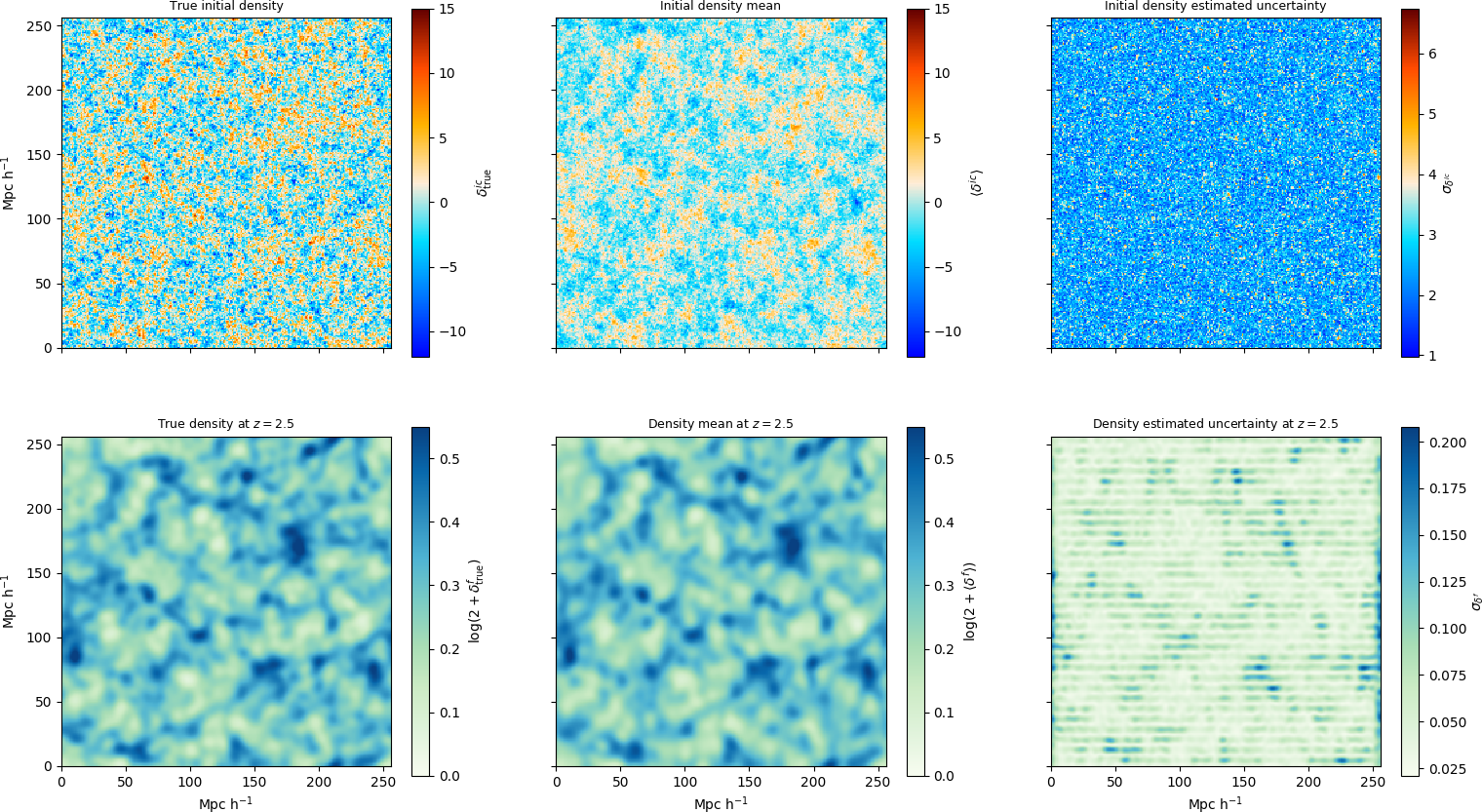}}
    \caption{Slices through ground truth initial (left upper panel), final density field (left lower panel), inferred ensemble mean initial (middle upper panel) and ensemble mean final (middle-lower panel) density field computed from 12600 MCMC samples. Comparison between these panels shows that the method recovers the structure of the true density field with high accuracy.
    We note that the algorithm infers the correct amplitudes of the density field. Right panels show standard deviations of inferred amplitudes of initial (upper right panel) and final density fields (lower right panel). The standard deviation of the final density field shows lower values at the position of the lines of sight. We note that the uncertainty of $\delta^\mathrm{f}$ presents a structure that correlates with the density field. Particularly, the variance is higher in high-density regions due to the saturation of the absorbed flux. In contrast, the standard deviation of the initial conditions are homogeneous and show no correlation with the initial density field due to the propagation of information from the final to the initial density field via the dynamical model.}
    \label{fig:panels}
    \end{figure*}
    
    \subsection{The warm-up phase of the sampler}
    \label{sec::warmup}

    In the large sample limit, any properly set up Markov chain is guaranteed to approach a stationary distribution that provides an unbiased estimate of the target distribution. While Markov chains are typically started from a place remote from the target distribution after a finite amount of transition steps, the chain acquires a stationary state. Once the chain is in a stationary state, we may start recording samples to perform statistical analyses of the inference problem. 
    To test when the Markov sampler has passed its initial warm-up phase, we follow a similar approach as described in previous works \citep{BORG,Foregrounds,Altair,PM,Robust}, by initializing the Markov chain with an over-dispersed state and trace the systematic drift of inferred quantities towards their preferred regions in parameter space. Specifically, we initialized the Markov chain with a random Gaussian cosmological density field scaled by a factor $10^{-3}$ and monitored the drift of corresponding posterior power-spectra during the initial warm-up phase. The results of this exercise are presented in Fig. \ref{fig:pk_burnin}. As can be seen, successive measurements of the posterior power-spectrum during the initial warm-up phase show a systematic drift of power-spectrum amplitudes towards their fiducial values. The sampler, therefore, correctly recovers the power of the initial density field and moves the chain towards regions of high probability in the parameter space. By the end of the warm-up phase, the sampler has found an unbiased representation of the matter distribution at all Fourier modes considered in this work. Starting the sampler from an over-dispersed state, therefore, provides us with an important diagnostics to test the validity of the sampling algorithm.
    
    \subsection{Statistical efficiency of sampler}
    \label{sec:efficiency}

   By design, subsequent samples in Markov chains are correlated. The statistical efficiency of an MCMC algorithm is determined by the number of independent samples that can be drawn from a chain of a given length. To estimate the statistical efficiency of the sampler, we estimate the correlation length of the astrophysical parameters $A$ and $\beta$ along the Markov chain. For a parameter $\theta$ the auto-correlation for samples with a given lag in the chain can be estimated as
    \begin{equation}
        C_n(\theta) = \frac{1}{N-n} \sum_{i=0}^{N-n} \frac{(\theta^i - \langle \theta\rangle)(\theta^{i+n} - \langle \theta \rangle)}{\mathrm{Var}(\theta)}
    \end{equation}
    where $n$ is the lag in MCMC samples, $\langle \theta\rangle$ is the mean and $\mathrm{Var}(\theta)$ is the variance. We typically determine the correlation length by estimating the lag $n_C$ at which the auto-correlation $C_n$ dropped below $0.1$. The number $n_C$ therefore present the number of transitions required to produce one independent sample.    
    The results of this test are presented in Fig. \ref{fig:meta_corr}. As can be seen, correlation lengths for parameters $A$ and $\beta$  amount to 640 and 830 samples, respectively. To significantly improve statistical efficiency, in the future, we use similar strategies to obtain a faster mixing of the chain as described in \citet{Altair}.

 \begin{figure}
    \centering
        {\includegraphics[width=\hsize,clip=true]{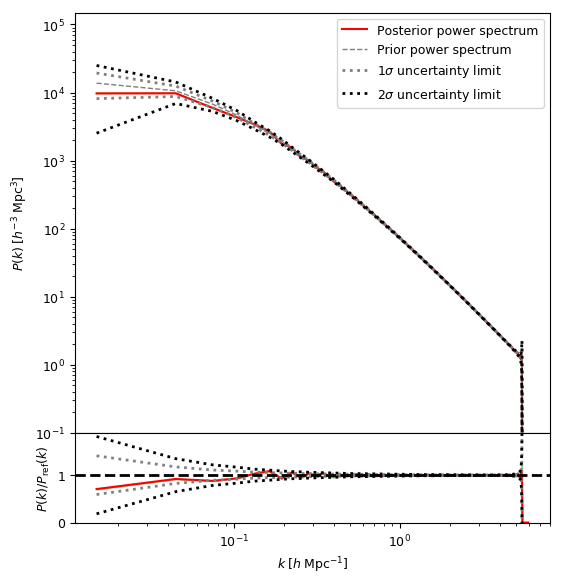}}
    \caption{Mean posterior matter power-spectrum. The mean and standard deviation of the matter power spectrum have been computed from 8060 density samples of the Markov chain obtained after the warm-up phase shown in Fig. \ref{fig:pk_burnin}. Although the standard deviation is plotted, it is too small to be seen, showing the stability of the posterior power-spectrum. The dashed line indicates the underlying power spectrum and the 1- and 2-$\sigma$ cosmic variance limit. The algorithm recovers the power-spectrum amplitudes within the $1-\sigma$ cosmic variance uncertainty limit throughout the entire range of Fourier modes considered in this work.}
    \label{fig:pk_mean}
    \end{figure}

    \begin{figure}
    \centering
        {\includegraphics[width=\hsize,clip=true]{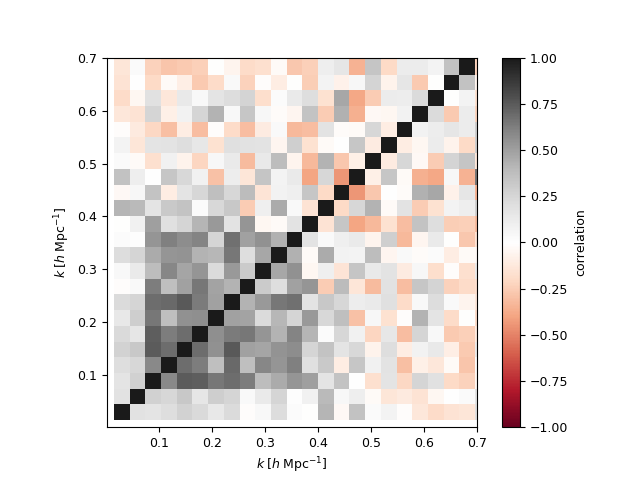}}
    \caption{Estimated correlation matrix of power spectrum amplitudes with the mean value, normalized using the variance of amplitudes of the power spectrum modes. The low off-diagonal contributions are a clear indication that our method correctly accounted and corrected for otherwise erroneous mode coupling, typically introduced by survey systematic effects and uncertainties. The expected correlations correspond to the lines of sight grid.}
    \label{fig:pk_corr}
    \end{figure}

    \subsection{Posterior predictions for quasar spectra}
 
   To test whether inferred density fields provide accurate explanations for the observations, we perform a simple posterior predictive test \citep[see e.g.][]{gelmanbda04}. Generally, posterior predictive tests provide good diagnostics about the adequacy of data models in explaining observations and identifying possible systematic problems with the inference.
   
    Here we use density fields and the astrophysical parameters $A$ and $\beta$, inferred by BORG, to predict expected quasar fluxes. If posterior predictions agree with actual observations within the uncertainty bounds, then the data model can be considered to be sufficient to analyse the data. In contrast, any misfits or systematic deviations would indicate a problem of the data model or the inference process.
    
    Specifically, we applied the FGPA model given in equation \ref{eq:FGPA} to inferred density fields:
    \begin{equation}
        (F_{pp})_{x} = \frac{1}{N}\sum_{i=0}^{N} \exp\Big[-A_i \big(1+(\delta^\mathrm{f}_i)_x\big)^{\beta_i}\Big]
    \end{equation}
    where $F_{pp}$ is the posterior-predicted flux, $i$ labels the samples and $x$ labels the volume elements.
    
    The result of this test is presented in Fig. \ref{fig:posterior_test}. As can be seen, the posterior predicted quasar spectrum nicely traces the data input within the observational $1\sigma$ uncertainty region. This demonstrates that the method correctly locates absorber positions and corresponding amplitudes of the underlying density field. 
    
    While Fig. \ref{fig:posterior_test} shows results only for a single line of sight, more generally, we also explored the flux errors for all lines of sight. The corresponding distribution of flux errors for posterior predictions is presented in
    Fig. \ref{fig:flux_error}. The distribution of flux errors corresponds to the distribution of pixel-noise in the spectra, demonstrating that our method is close to the theoretical optimum. 
    This indicates that our method exhibits good control of the data model, including the handling of nuisance parameters and uncertainties.
    
    Since many studies of the Ly-$\alpha$ forest are based on the power spectrum of the flux, we tested the flux power spectrum of the posterior-predicted fluxes. The results of this test are presented in Fig. \ref{fig:pk_flux}, showing that the power spectrum of the recovered spectra matches the true flux power spectrum very accurately. We note that we computed the power spectrum of the flux from the noiseless data as the posterior-predicted fluxes do not contain noise.

    \begin{figure}
    \centering
        {\includegraphics[width=\hsize,clip=true]{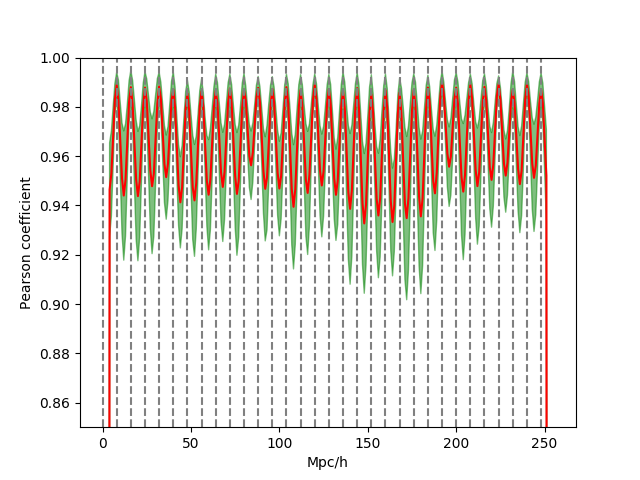}}
    \caption{Pearson coefficient of the inferred and true density field as a function of the position across lines of sight. The dashed lines indicate the positions of the lines of sight. The Pearson coefficient is $>0.9$ at any location in the density field. This, combined with Fig. \ref{fig:panels} showing that the algorithm recovers the correct amplitudes,  indicates that the algorithm can interpolate the information between lines of sight and correctly recover the structures in unobserved regions.}
    \label{fig:pearson}
    \end{figure}
    
    \subsection{Isotropy of the velocity field}
    One-dimensional Ly-$\alpha$ forest data introduce a particular geometry of the data into the inference problem. In particular, data are only available along one-dimensional lines of sight. In this work, we are interested in recovering the three-dimensional density field from Ly-$\alpha$ forest data by fitting a dynamical model. This model describes the formation of structures by displacing matter from its initial conditions to their final Eulerian positions. As such, we have found, that the large-scale velocity field is particularly sensitive to systematic effects introduced by the data, in particular, the survey geometry. Since the distribution of lines of sight defines a preferred axis, we thus test the isotropy of the velocity field to ensure that the algorithm correctly recovered a physically plausible three-dimensional velocity and corresponding density fields. The results of this test are shown in Figure \ref{fig:velocities}, showing that there is no preferred component of the three-dimensional velocity field. All velocity components have a similar distribution indicating that the algorithm correctly accounted and corrected for the geometry of the survey. This result is further supported by the recovery of an isotropic primordial power-spectrum as described in section \ref{sec:results_Pk_lya}.

\section{Analysing the LSS in Ly-$\alpha$ forest data}
\label{sect:lya_results}
In this section, we present the results of applying our algorithm to simulated Ly-$\alpha$ forest data. We show that our method infers physically plausible and unbiased density fields and corresponding power-spectra at all scales considered in this work. We further demonstrate that the algorithm accurately extrapolates information into unobserved regions between lines of sight. We further illustrate the performance of the algorithm by recovering mass and velocity profiles of clusters and voids. 
    
    \subsection{Inference of matter density fields at high-redshift}
    
    As discussed above, the BORG algorithm performs a full-scale Bayesian analysis of the cosmological large-scale structure in Ly-$\alpha$ forest data. This is achieved by using the physical forward modelling approach to fit a dynamical model of structure growth to the data. As a result, we simultaneously obtain inferences of the primordial field of fluctuations from which structures formed, the non-linear spatial matter distribution at a redshift of $z=2.5$ and corresponding velocity fields. More specifically, the BORG algorithm provides us with an ensemble of realizations of these fields drawn from the corresponding large-scale structure posterior distribution discussed in Appendix \ref{app:posterior_lya}. This ensemble of realizations from the Markov Chain enables us to estimate any desired statistical summary and quantify corresponding uncertainties. 
    
    As an example, in Figure \ref{fig:panels}, we illustrate ensemble mean and variances for primordial and final density fields. Specifically, Figure \ref{fig:panels}  shows slices through the true density, and the ensemble mean and variances of inferred three-dimensional density fields, computed from 12600 samples. 
    A first visual comparison between ground truth and the inferred ensemble mean final density fields in the lower panels of figure \ref{fig:panels} illustrates that the algorithm correctly recovered the three-dimensional large-scale structure from Ly-$\alpha$ forest data. On first sight, both fields are visually almost indistinguishable, indicating the high quality of the inference. The lower right panel of figure \ref{fig:panels} shows the corresponding standard deviations for density amplitudes for respective volume elements as estimated from the Markov Chain. It can be seen that the estimated density standard deviations correlate with the inferred density field. This is expected for a non-linear data model, which couples signal and noise. In particular, one can observe that uncertainties are lowest for under-dense regions where Ly-$\alpha$ forest data provide high signal-to-noise, as quasar light is simply transmitted by structures. In contrast, one can observe the highest uncertainties for over-dense structures. Since over-dense structures absorb quasar light, this decreases the signal in these regions: the absorption is saturated at high densities. Even more, if structures are sufficiently dense, then they absorb all quasar light and the absorption becomes saturated. Once light absorption is saturated, the data provide no further information about the actual density amplitude of the absorber and data only provide information on a minimally lowest density threshold required to explain the observations. Figure \ref{fig:panels} therefore, clearly demonstrates that our method correctly accounts for and quantifies uncertainties inherent to Ly-$\alpha$ forest observations.
    It is interesting to remark, that while observations of galaxy clustering are most informative in high-density regions, the Ly-$\alpha$ forest is particularly informative for under-dense regions and therefore provides complementary information to galaxy surveys.
    Figure \ref{fig:panels} demonstrates that, in addition to inferring the correct filamentary structure, our method clearly infers the correct amplitudes of the density field. Our results can also be compared to Fig. 2 in \cite{Horowitz19}, which presents a systematic underestimation of the density amplitude.

        \begin{figure*}[ht] 
  \begin{subfigure}[b]{0.5\linewidth}
    \centering
    \includegraphics[width=\linewidth]{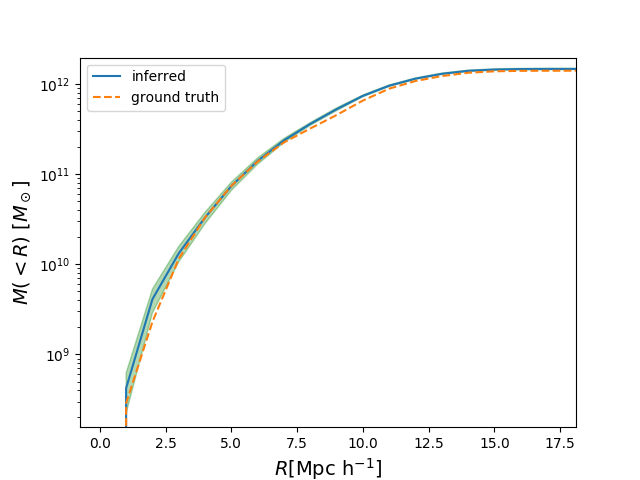} 
    \vspace{4ex}
  \end{subfigure}
    \begin{subfigure}[b]{0.5\linewidth}
    \centering
    \includegraphics[width=\linewidth]{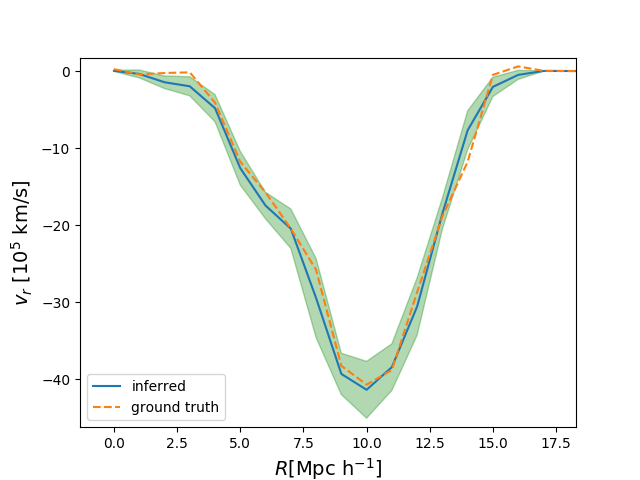} 
    \label{fig7:b} 
    \vspace{4ex}
  \end{subfigure} 
  \begin{subfigure}[b]{0.5\linewidth}
    \centering
    \includegraphics[width=\linewidth]{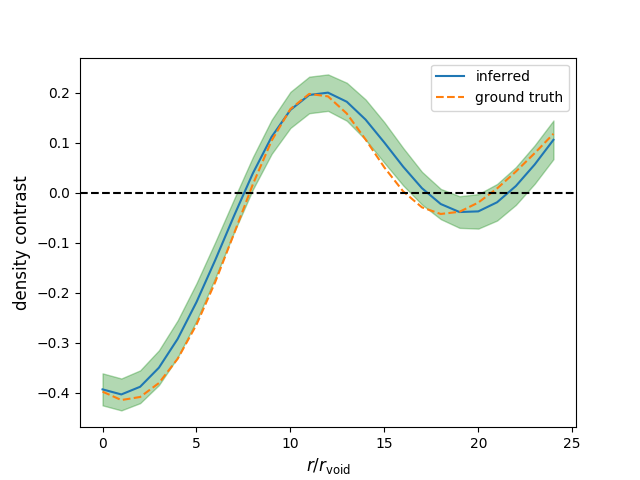} 
  \end{subfigure}
  \begin{subfigure}[b]{0.5\linewidth}
    \centering
    \includegraphics[width=\linewidth]{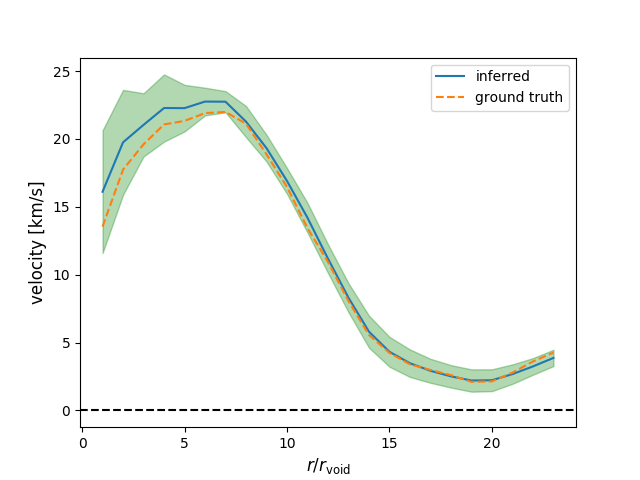} 
  \end{subfigure} 
  \caption{Top: Cluster mass (left) and radial velocity (right) profiles from the inferred ensemble mean density field and the simulation.  These profiles are computed by averaging over the mass and velocity profiles measured on 60 different density samples. The shaded regions show the standard deviation between these samples. The structured is traversed by 4 lines of sight with a Gaussian noise distribution of $\sigma=0.03$. The algorithm recovers the correct mass profile from the simulation. This demonstrates that the method provides unbiased mass estimates, correcting for the astrophysical bias of the IGM. Bottom: Spherically averaged density (left) and velocity (right) profiles of a void, obtained by averaging the void profile measured in 60 density samples. The standard deviation between the samples is shown as the shaded region. This void is traversed by 6 lines of sight with a Gaussian noise distribution of $\sigma=0.03$. These plots show that the method recovers the underlying density and velocity profiles of the simulation.}
        \label{fig:cluster_mass}
    \end{figure*}
    
   \subsection{Analysing posterior power-spectra}
    \label{sec:results_Pk_lya}
    
    As demonstrated in the previous section, the algorithm provides accurate reconstructions of final density fields. To provide more quantitative tests and to estimate whether inferred density fields are physically plausible, we perform analyses of posterior power-spectra measured from inferred density fields. 
    
    Reconstructing three-dimensional density fields from one-dimensional Ly-$\alpha$ data is technically challenging.
    For example, \cite{Kitaura12} used a Gibbs sampling approach to sample the large- and small-scales of the density field separately with a lognormal prior for the evolved density field. This approach inferred correct power-spectrum amplitudes at large scales $k<0.1$ h Mpc$^{-1}$ but obtained erroneous excess power at smaller scales, which was attributed to the inadequacy of the log-normal approximation or complex correlations in the data \citep{Kitaura12}. \cite{Horowitz19} used an optimization approach to fit a dynamical forward model to the data, but the method obtains power-spectra that severely underestimate the power of density amplitudes. 
    
    As we aim to optimally extract cosmologically relevant information from Ly-$\alpha$ forest data, we are particularly interested in recovering physically plausible density fields from observations. Erroneous power in inferred power-spectra is often a sign of untreated or unknown survey systematics and indicates a break down of the assumptions underlying the data model or the inference approach.  In this work, we use a sophisticated MCMC framework to marginalize out nuisance parameters and quantify uncertainties inherent to the observations. As already indicated in section \ref{sec::warmup}, our approach is able to identify the correct power distribution of density amplitudes from observations. To further quantify this result, we estimate the mean and variance of posterior power-spectra measured from the ensemble of Markov samples. The results are presented in Fig. \ref{fig:pk_mean}. As can be seen, our method recovers the correct fiducial cosmological power-spectrum within the 1-$\sigma$ cosmic variance uncertainty at all Fourier modes considered in this work. The unbiased recovery of the power-spectrum clearly indicates that the method correctly accounted for uncertainties and systematic effects of the data.
    
    Systematic effects, such as survey geometries, often introduce spurious correlations between power-spectrum modes, when not accounted for properly. This is of particular relevance for Ly-$\alpha$ data, which introduces a particular survey geometry through the set of one-dimensional lines of sight. To test for residual correlations between Fourier modes, we estimated the covariance matrix of power-spectrum amplitudes from our ensemble of Markov samples. As shown in Fig. \ref{fig:pk_corr} this covariance matrix shows a clear diagonal structure with the expected correlations at $k\sim 0.5-0.7$ h Mpc$^{-1}$ due to the lines of sight grid. This test shows that our method correctly accounted for survey geometries and other systematic effects that could introduce erroneous mode coupling.
    
    In summary, these tests demonstrate that our method is capable of inferring physically plausible matter density fields with correct power distribution from noisy Ly-$\alpha$ data. 
    
     \begin{figure}
    \centering
        {\includegraphics[width=\hsize,clip=true]{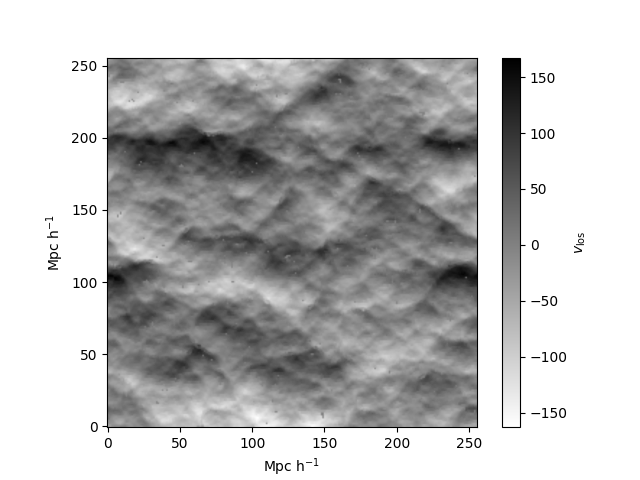}}
    \caption{Projection of the velocity along lines of sight, which are parallel to the $x$-axis of the plot.}
    \label{fig:vel_vector}
    \end{figure}

    \begin{figure}[t]
    \centering
        {\includegraphics[width=\hsize,clip=true]{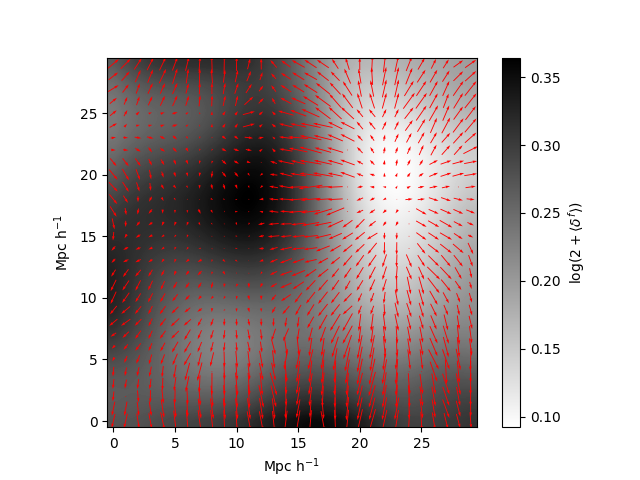}}
    \caption{Zoom-in on the density field. The vector field shows the velocities on top of the inferred density field, showing matter flowing out of the void and falling into the gravitational potential of the cluster. Therefore, the method provides consistent velocity and density fields.}
    \label{fig:structure}
    \end{figure}

    \subsection{Recovering information between lines of sight}

    The previous section describes that inferred density fields are physically plausible realizations of the matter field underlying the Ly-$\alpha$ observations. Here we want to quantify to what accuracy the method recovers the underlying ground truth density field. A particular challenge is to correctly recover the density field in between one-dimensional lines of sight. As illustrated above, our method is capable of recovering the three-dimensional density field from a set of one-dimensional quasar spectra using the Bayesian physical forward modelling approach. In particular, we determine the performance of our algorithm to recover the density field from data by estimating the Pearson correlation coefficient (see Appendix \ref{app:pearson}) between the inferred and ground truth density fields. We estimate the Pearson correlation coefficient as a function of the position on the $x$-axis, which permits us to track the correlations on and in-between observed lines of sight. 
    Fig. \ref{fig:pearson} shows the Pearson coefficient averaged over 300 samples of the chain together with its 1-$\sigma$ credibility interval. It can be clearly seen that inferred density fields correlate with the ground truth density typically to more than 98\% along the lines of sight. But even in-between lines of sight the correlation is on average larger than 90\%. These high Pearson correlations demonstrate the capability of our method to recover the cosmic large-scale structure in unobserved regions between observed lines of sight.
    
    \subsection{Cluster and void profiles}

        \label{sec:meta_results_lya}
    \begin{figure*}[t]
        \centering
            {\includegraphics[width=0.7\hsize,clip=true]{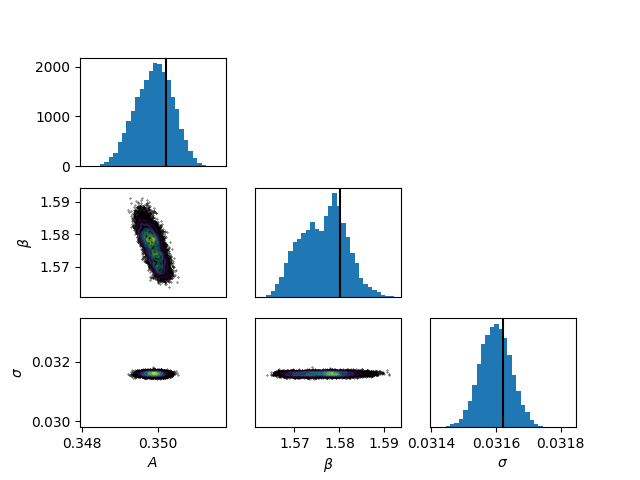}}
        \caption{Corner plot for the unknown parameters $A$ and $\beta$ of the FGPA model and the standard deviation of the Gaussian noise $\sigma$. As indicated in the plot, different panels show different marginal posterior distributions for respective parameters. Black solid lines in uni-variate marginal distributions indicate the true fiducial parameter values. It can be seen that the method correctly infers the underlying parameters and quantifies corresponding uncertainties. It is interesting to remark, that the astrophysical parameters $A$ and $\beta$ show no a posteriori correlations with the noise standard deviation  $\sigma$.}\label{fig:meta}
    \end{figure*}

    Above we showed that the BORG algorithm recovers physically plausible density fields from Ly-$\alpha$ data that agree with the expected statistical behaviour of a $\Lambda$CDM density field. Here we want to test if the algorithm also correctly recovers the properties of individual cosmic structures at their respective locations in the three-dimensional volume. To illustrate this fact, we study the properties of inferred clusters and voids. In particular, we show that the algorithm correctly recovers mass density and velocity profiles of cosmic structures. 

    To test the properties of an inferred cluster, we randomly chose a peak in the final density field. Then we determined the mass and velocity profiles in spherical shells around the identified centre of the cluster. In particular, we estimated the cumulative radial mass profiles as:
    \begin{equation}
        M(<r) = \frac{1}{N} \sum_{i=0}^N m_p \, K^i_\mathrm{part}(<r)  
    \end{equation}
    where $r$ is the radial distance from the cluster centre, $i$ labels the Markov samples, $N$ is the total number of samples and  $K^i_\mathrm{part}(<r)$ is the number of simulation particles of Markov sample $i$ inside a sphere of radius $r$ around the cluster centre. Finally, $m_p=287\times 10^6$ M$_\odot$  is the particle mass of the simulation, which is obtained as 
    \begin{equation}
        m_p = \frac{\bar{\rho}  V_\mathrm{box}}{N_\mathrm{part}^\mathrm{total}}
        \label{eq:particle_mass}
    \end{equation}
    where $\bar{\rho}$ is the mean cosmic density, $V_\mathrm{box}$ is the volume of the inferred density field and $N_\mathrm{part}^\mathrm{total}$ is the total number of particles used in the simulation.  Fig. \ref{fig:cluster_mass} shows the cluster mass profile measured from the inferred and true density fields. The inferred mass profile shown here is the average over 60 samples. Fig. \ref{fig:cluster_mass} shows that our method provides unbiased mass estimates of cluster mass profiles, becoming an alternative to measuring cluster masses, complementary to weak lensing or X-ray observations. 

    The method also provides inferred density and velocity profiles of voids. Fig. \ref{fig:cluster_mass} shows the density and velocity profiles measured from the inferred and true density fields. The void is defined spherically and the velocity and density profiles are spherically averaged: 
    \begin{eqnarray}
    \rho(r)&=& \frac{1}{N} \sum_{i=0}^N \frac{3 m_p  K^i_\mathrm{part}(<r) }{4\,\pi r^3} \\
    v(r) &=& \frac{1}{N} \sum_{i=0}^N \frac{1}{K^i_\mathrm{part}(<r) }
    \sum_{p}^{K^i_\mathrm{part}(<r) } v_p 
    \end{eqnarray}
    where $i$ runs over the samples, $N$ is the total number of samples, $r$ is the radius of the sphere centred in the volume element with the lowest density, $m_p$ is the particle mass described in eq. \ref{eq:particle_mass}, $K^i_\mathrm{part}(<r)$ is the number of particles inside the sphere and $v_p$ are the velocities of the particles provided by the dynamical model. The profiles shown in Fig. \ref{fig:cluster_mass} are obtained from averaging over 60 samples. Voids constitute the dominant volume fraction of the Universe. Since the effect of matter in voids is mitigated, they are ideal to study the diffuse components of the Universe such as dark energy \citep{VoidDE08,VoidDE10,VoidsGL,VoidsDE12} and gravity \citep{VoidGR11,VoidGR13,VoidGR13b,VoidGR16}. Voids are also interesting tools to constrain the neutrino mass \citep{VoidNeutrino15,NeutrinoVoid,VoidNeutrino19,VoidNeutrino19b}: the neutrino free-streaming length falls within the range of typical void size. Therefore, the void profiles obtained with this framework provide an alternative tool to constrain the neutrino mass.

    \subsection{Velocity field and structure formation}

    The dynamical model employed in BORG allows us to naturally infer the velocity field since it derives from the initial perturbations. The velocity information allows discriminating between peculiar velocities and the Hubble flow. The combination of velocity and density fields can provide significant information on the formation of structures and galaxies. 
    
    Our method provides velocity fields at $z=2.5$, where this kind of data is usually hard to obtain. Fig. \ref{fig:vel_vector} shows the velocity along the line of sight. The line-of-sight velocity provides a tool to measure the kinetic Sunyaev–Zeldovich effect \citep{SZ72,SZ80,SZ3} by cross-correlating the velocity field with the CMB and study the expansion of the Universe.
    
    While the hierarchical structure formation model has been tested in the nearby Universe \citep{Hierarchical82,Hierarchical84,Hierarchical85}, our method provides a framework to test it at high redshift. The method provides consistent velocity and density fields, as shown in Fig. \ref{fig:structure}, which shows a zoom-in on the inferred mean density field and the corresponding velocity components $(v_x,v_y)$. Therefore, we can test the hierarchical structure formation model at high redshift by combining the density and velocity information and testing the predictions of the method with observations. Additionally, these density and velocity fields can provide some insights into galaxy formation and evolution by studying the effect of large-scale structures in galaxy populations \citep{AGNpaper}. Therefore, this method for the Ly-$\alpha$ forest allows extending investigations of the nature of galaxies or AGN to the high-redshift Universe.

    \begin{figure*}
        \centering
            {\includegraphics[width=\hsize,clip=true]{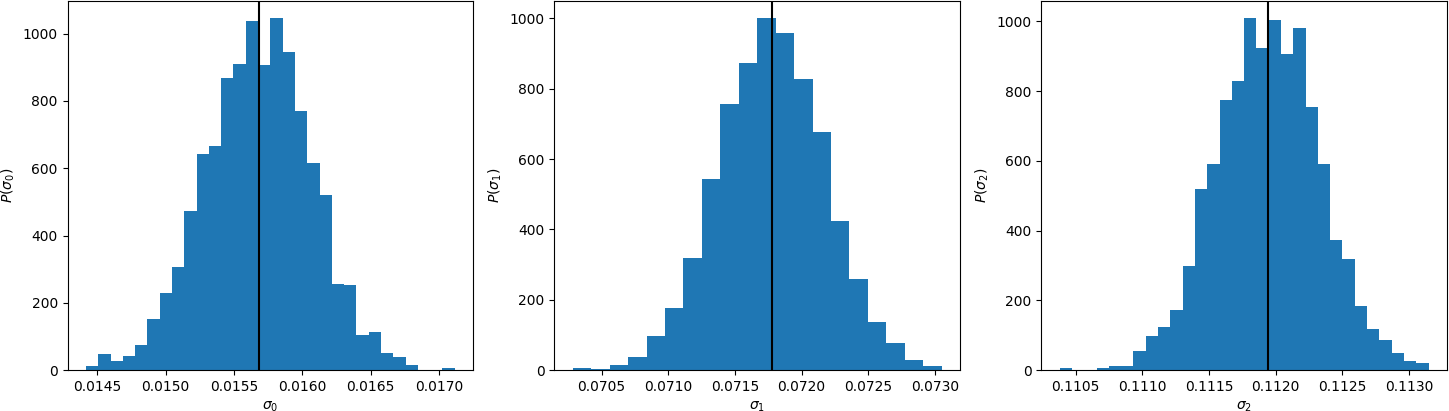}}
        \caption{Posterior distribution of the noise standard deviations for three lines of sight. The method samples the individual $\sigma$ of each line of sight, recovering the input value (solid black line).}\label{fig:meta_individual}
    \end{figure*}
    
\begin{figure}
        \centering
            {\includegraphics[width=\hsize,clip=true]{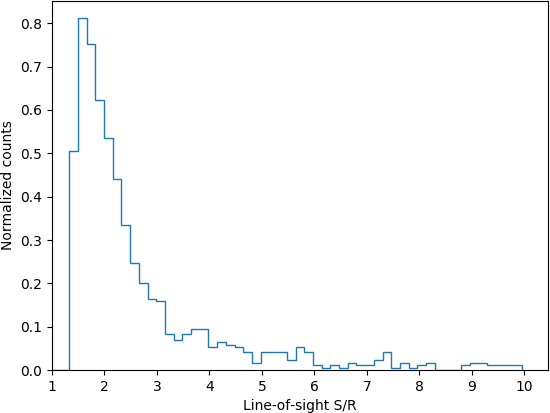}}
        \caption{Distribution of signal-to-noise (S/N) per pixel in different lines of sight. This distribution of S/N follows a power-law, mimicking the CLAMATO survey \citep{CLAMATO_SNR}. The S/N is constant along each line of sight but varies among different lines of sight.}\label{fig:snr}
    \end{figure}
    
    \subsection{The astrophysical parameters}

    Current efforts of the science community aim at constraining the astrophysics of the intergalactic medium \citep[see e.g.][]{Lee12,InvertedIGM17,IGM-17Cont} since the thermal history of the IGM holds the key to understanding hydrogen reionization at $z>6$ \citep{Lee12}. Particularly, analysis of the Ly-$\alpha$ forest attempt to measure the parameters $T_0$ and $\gamma$. The parameter $T_0$ is related to the spectral shapes of the ultraviolet background. Therefore, it contains information on the intensity of the ionizing background, which is relevant to understand the nature of the first luminous sources and their impact on the subsequent generation of galaxies \citep{ReionMotivation}. The parameter $\gamma$ defines the temperature-density relation indicating whether over-dense regions are hotter than the under-dense regions ($\gamma > 1$) or vice versa ($\gamma < 1$). While \citet{InvertedIGM07,InvertedIGM08,InvertedIGM09,InvertedIGM17} found evidence for $\gamma < 1$, \citet{NeutralIGM12,NeutralIGM12b} found $\gamma \approx 1$. Other approaches \citep{PositiveIGM00,PositiveIGM00b,PositiveIGM01,PositiveIGM10,PositiveIGM12,PositiveIGM14} found $\gamma > 1$, which is more in agreement with the theoretically predicted $\gamma \approx 1.6$ for a post-reionization IGM \citep{LyPhysics}. However, the value of $\gamma$ is still an open debate since the different values are obtained from different kind of data, affected by systematic uncertainties in a different way \citep{IGM-17Cont}.  
    
    The algorithm presented in this work infers the astrophysical parameters of the FGPA, which are directly related to $T_0$ and $\gamma$. Fig. \ref{fig:meta} shows the posterior distribution and correlations of these parameters and the noise variance $\sigma$, obtained from 6800 samples. 
    This shows that the method recovers the true value of the posterior distribution of the FGPA parameters, corresponding to $T_0 = (206.5 \pm 0.1) 10^3\Gamma^{-1/0.7} $K and $\gamma=1.614\pm0.004$, where the uncertainty is derived from the standard deviation of $A$ and $\beta$. While the more accurate measurements of $\gamma$ have an uncertainty above 7\% \citep{PositiveIGM12,IGM-17Cont}, this test shows that our method can constrain $\gamma$ with an uncertainty below $1\%$. The reason for this tight constraint of $\gamma$ is probably the fact that higher-order statistics of the density field can break parameter degeneracy \citep{Fabian}. Although constraining $\gamma$ from real Ly-$\alpha$ forest data require to model the continuum flux of the quasar, our method holds the promise to shed new light on the temperature-density debate. 
    
    Note in Fig. \ref{fig:meta} that parameters $A$ and $\beta$  do not show a strong correlation with the noise variance $\sigma$, which is jointly sampled. It is also worth noticing that $\beta$ presents a bimodal distribution. Optimization methods \citep[see e.g.][]{Horowitz19}
    can only explore local extremes of the distribution, becoming sub-optimal to explore multi-modal distributions. However, our MCMC approach can characterize the bimodal distribution and provide the corresponding uncertainties, which are necessary to interpret the data correctly. Therefore, our algorithm provides a framework to constrain the astrophysics of the intergalactic medium jointly with the density field. To the knowledge of the authors, this work presents the first approach that attempts to jointly quantify the astrophysical parameters and the 3d cosmic structure.

    \begin{figure*}
        \centering
            {\includegraphics[width=0.95\hsize,clip=true]{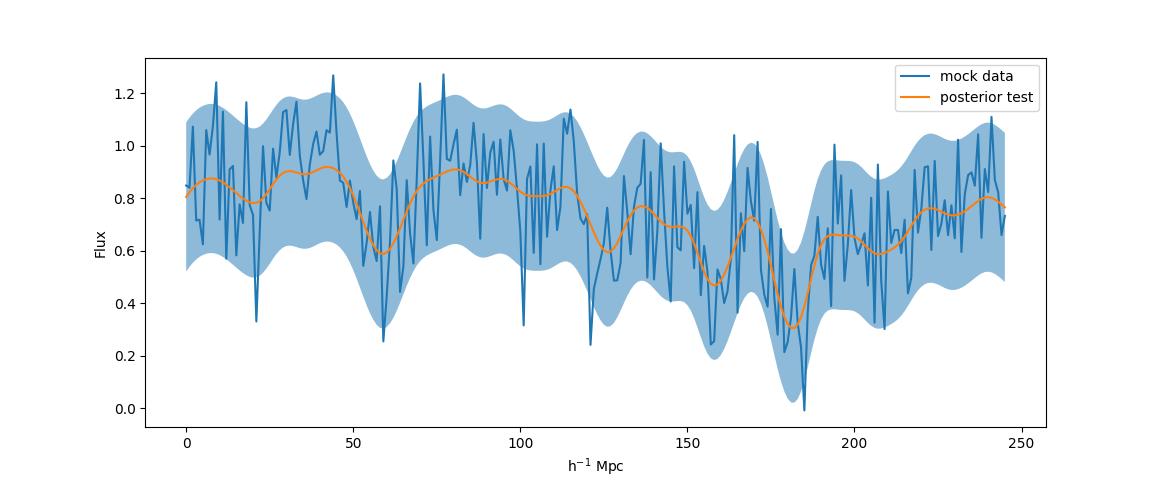}}
         {\includegraphics[width=0.95\hsize,clip=true]{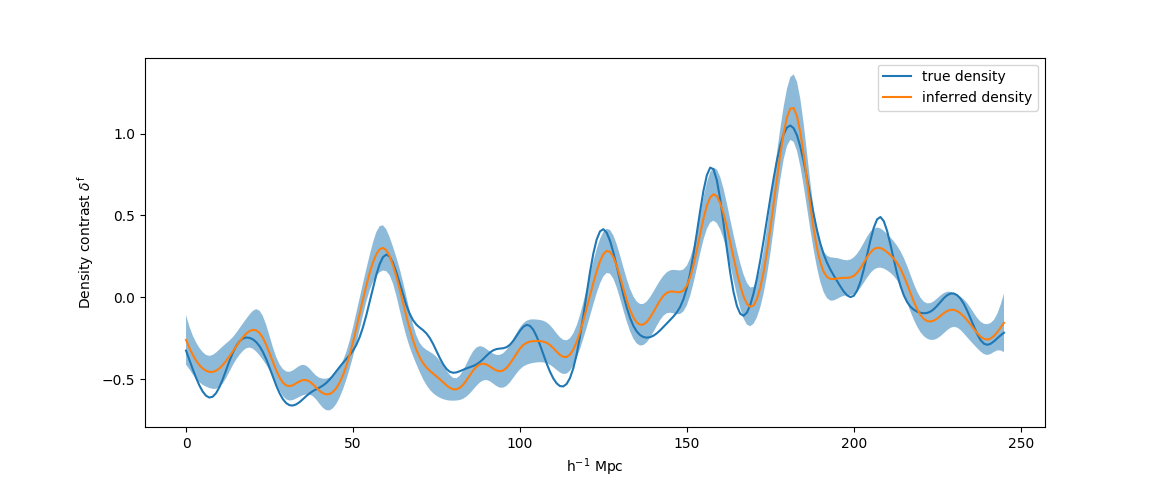}}
        \caption{Top panel: Posterior predictive flux for a spectrum with $\mathrm{S/N} = 2$. The posterior predicted flux is computed from the ensemble mean density field, shown in the lower panel (orange line). The blue shaded region indicates the 1-$\sigma$ region, corresponding to the standard deviation of the noise in this line of sight.  Our method recovers the transmitted flux fraction correctly, confirming that the data model can accurately account for the observations also in the low S/N regime. Bottom panel: Comparison of the inferred ensemble mean density field along the line of sight to the ground truth. The density field has been smoothed with a Gaussian smoothing kernel of $\sigma=0.5 h^{-1}$~Mpc to simulate the difference between dark matter and gas density fields \citep{Peirani14}. This smoothing kernel is applied to the dark matter density field before generating the mock data.
        \label{fig:ppt_low_snr}}
    \end{figure*}

\begin{figure}
        \centering
            {\includegraphics[width=\hsize,clip=true]{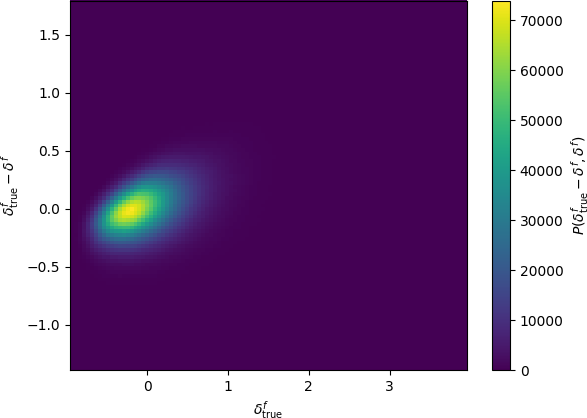}}
        \caption{Probability distribution of the difference between the true density field and the ensemble mean density. The algorithm correctly recovers the unbiased density field from the Ly-$\alpha$ forest data with low S/N and individual noise distribution for each spectra.}\label{fig:scatter}
\end{figure}

\section{Study with low signal-to-noise data}
\label{sec:snr}

The previous tests and analysis were performed using simulated data with a noise distribution that was the same for all lines of sight. However, in real data, different lines of sight present different signal-to-noise ratio (S/N), which depend on the integration time of the spectrum. For this reason, this section presents an analysis with mock data with different S/N for each spectrum. The distribution of lines of sight is the same as described in Sec. \ref{sect:mock_data} but the Gaussian pixel-noise is characterized by a different S/N in each line of sight. Therefore, the S/N is constant along one line of sight but varies among lines of sight, following a power-law as described in \cite{CLAMATO_SNR} and \cite{Horowitz19}.

To simulate realistic pixel-noise, we draw S/N values for each line of sight according to
\begin{equation}
\frac{\mathrm{d} n_\mathrm{los}}{\mathrm{d}\mathrm{S/N}} \propto \mathrm{S/N}^\alpha    
\end{equation}
with $\alpha = 2.7$ \citep{Horowitz19}. The distribution of S/N is shown in Fig. \ref{fig:snr}, which can be compared to Fig. 2 in \cite{CLAMATO_SNR}. Mimicking the CLAMATO survey, we limit the values of possible S/N to $\mathrm{S/N}_\mathrm{min} = 1.4$ and $\mathrm{S/N}_\mathrm{max} = 10$ \citep{Horowitz19}. The Gaussian pixel-noise is then drawn according to the S/N of each line of sight. The individual $\sigma$ of the noise distribution for each line of sight is sampled jointly with the astrophysical parameters $A$ and $\beta$. Figure \ref{fig:meta_individual} shows that the method can recover the individual noise distribution for different spectra.

Figure \ref{fig:ppt_low_snr} shows a posterior predicted test for a low S/N spectrum (S/N=2), illustrating the performance of the method in a low S/N regime. This figure shows that the algorithm can predict the input flux and recover the matter density along the line of sight in the 1-$\sigma$ uncertainty limits.

Figure \ref{fig:scatter} shows the probability distribution of the difference between the true density field and the ensemble mean density, demonstrating that the algorithm recovers the unbiased matter density. This test, therefore, shows that our method can be successfully applied to low S/N data, providing the unbiased dark matter density field, jointly inferred with the astrophysical parameters and the characterization of the noise distribution.

\section{Summary and discussion}
\label{conclusions}

Observations of the late time cosmic large-scale structure can provide information on fundamental physics driving the dynamic evolution of our Universe only if we manage to accurately account for non-linear data models and systematic effects on the data, and if we connect theory to observations. A detailed physical understanding of our universe, therefore, requires linking the current increase of high-quality cosmological data with the development of novel analysis methods capable of making the most of such observations. Traditional analyses of cosmic structures are limited to exploiting only one- and two-point statistics, but ignore significant information encoded in higher-order statistics associated with the filamentary features of the late time cosmic matter distribution. To fully account for the information content of the cosmic large-scale structure, we need to follow a field-based approach to extract the entire three-dimensional matter distribution from observations. In this work we were particularly interested in studying the cosmic matter distributions at redshifts $z>2$ with one-dimensional Lyman-$\alpha$ forest observations.

To extract the full three-dimensional information content from the Ly-$\alpha$ forest, we have proposed a Bayesian physical forward modelling approach to fit a numerical model of gravitational structure formation to data. Building upon the previously presented BORG algorithm, our fully probabilistic framework infers the three-dimensional matter density field and its dynamics from the Lyman-$\alpha$ forest. The method is a considerable improvement over previous approaches \citep[e.g.][]{Kitaura12, Horowitz19} as it infers unbiased dark matter fields from the Ly-$\alpha$ forest for the first time. While the previous approaches failed to recover the matter-power spectrum, our method provides the unbiased matter power spectrum at all range of Fourier modes. In particular, the approach followed by \citet{Horowitz19} produces an underestimation of the power spectrum, corresponding to too low amplitudes of the inferred density field. We have shown that our method recovers the unbiased dark matter distribution and physically plausible density and velocity fields from Lyman-$\alpha$ data. 

To make such inferences feasible, we used a likelihood function based on the fluctuating Gunn-Peterson approximation (FGPA). In addition to modelling the dynamical evolution and spatial distribution of cosmic matter, our approach also accounts for systematic effects arising from the astrophysical properties of the intergalactic medium by self-consistently marginalizing out corresponding nuisance parameters of the FGPA model. We have shown that this approach can provide tight constraints on the astrophysical parameters  $T_0$ and $\gamma$, which provide significant information about the reionization history of the Universe and the early luminous sources. 

Our hierarchical Bayesian framework encodes non-linear dynamical models. This work is based on the Lagrangian Perturbation Theory (LPT). As demonstrated by hydrodynamical simulations, the Ly-$\alpha$ forest arises from regions where the density of matter is within a factor ten of the cosmic mean density, and thus is still close to the linear regime of gravitational collapse \citep{Peirani14,Sorini16}. 

We tested our approach using realistic mock observations emulating the CLAMATO survey to infer the dark matter density and the velocity field at redshift $z = 2.5$ with a resolution of 1$h^{-1}$~Mpc. These tests demonstrate that our method recovers unbiased reconstructions of the non-linear spatial matter distribution and its power-spectrum from observations of the Ly-$\alpha$ forest. The inferred mean density field presents a high correlation with the true density at any location. This indicates that our method can interpolate the information between lines of sight. While previous approaches tried to address this challenge with polynomial smoothing \citep{Cisewski14} or Wiener filtering \citep{Ozbek16, Stark15}, the dynamical model in our method interpolates the information by accounting for the high-order statistics of the density field. 

The dynamical model naturally infers velocity fields jointly with the density field. Therefore, our method provides velocity fields at $z>2$, where this kind of data is usually hard to obtain. From the velocity fields, we can derive the velocities along the lines of sight or radial velocities. These velocity fields can be cross-correlated with the CMB to detect the kinetic Sunyaev–Zeldovich effect. Additionally, the consistent velocity and density fields provide a framework to test hierarchical structure formation and galaxy formation models at high redshift. 

We have shown that the method provides accurate mass and velocity profiles for cosmic structures, such as voids and clusters. Therefore, this method provides an alternative to measuring cluster masses, complementary to X-ray and lensing measurements. The method also provides the velocity and density profile of voids. Since the non-linear effects are mitigated in voids, they are sensitive to the diffuse components of the Universe such as dark energy. Therefore, the velocity and density profiles of voids can be used to discriminate between homogeneous dark energy and modifications of gravity. 

Since the Ly-$\alpha$ probes the matter distribution down to few megaparsecs, it is sensitive to neutrino masses. This high resolution and the void profiles provided by our algorithm can constrain neutrino masses. Additionally, this high resolution of the Ly-$\alpha$ forest allows us to infer the density field at the 1 Mpc scale. By applying new techniques compatible with  \cite{Altair}, this high-resolution density field can provide tight constraints of the cosmological parameters by studying structure geometry.

Summarizing, our method clearly demonstrates the feasibility of detailed and physically plausible inferences of three-dimensional large-scale structures at high redshift from the Ly-$\alpha$ forest observations. The proposed approach, therefore, opens a new window to study cosmology and structure formation at high redshifts.

\section*{Acknowledgements}
This work has made use of the C2PAP cluster at the Leibniz-Rechenzentrum.
NP is supported by the DFG cluster of excellence ``Origin and Structure of the Universe''\footnote{www.universe-cluster.de}.  GL acknowledges financial support from the ILP LABEX (under reference ANR-10-LABX-63) and from ``Programme National de Cosmologie et Galaxies'' (PNCG) of CNRS/INSU, France, and the ANR BIG4, under reference ANR-16-CE23-0002. This work was carried out within the Aquila Consortium\footnote{\url{https://aquila-consortium.org}}.

\bibliographystyle{aa} 
\bibliography{biblio} 

\appendix

\section{Projector along the line of sight}
\label{app:projector}

The Ly-$\alpha$ forest is measured along lines of sight to background sources. The volume elements intersected by the line of sight are identified from the wavelengths in the spectrum: first, the redshift is computed from the wavelength
\begin{equation}
z = \frac{\lambda - 1}{\lambda_0}
\end{equation}
where $\lambda_0 = 1216$ \AA \ corresponding to the wavelength of the Ly-$\alpha$ transition. The redshift is then used to compute the comoving distance from tabulated values of $d_\mathrm{com}\left(z\right)$ and finally, this is transformed into the regular grid coordinates as
\begin{eqnarray}
x &=& d_\mathrm{com}\left(z\right) \cos\left(\mathrm{\xi}\right)\cos\left(\mathrm{\alpha}\right), \\
y &=& d_\mathrm{com}\left(z\right) \cos\left(\mathrm{\xi}\right)\sin\left(\mathrm{\alpha}\right), \\
z &=& d_\mathrm{com}\left(z\right) \sin\left(\mathrm{\xi}\right),
\label{position_sources_borg}
\end{eqnarray}
with $\xi$ being the declination and $\alpha$ being the right ascension of the background source.

\section{Method}
\label{method}
In this section, we present a more detailed description of our algorithm. We describe the data model, likelihood, and prior. We briefly discuss how the high-dimensional joint posterior distribution can be explored via a multiple-block sampling scheme. 

\subsection{Prior}
    
BORG includes a physical density prior involving a model for structure formation. Therefore, it translates the inference of the density field into the inference of the initial conditions, which obey Gaussian statistics. The link between the initial and evolved density fields is given by the deterministic structure formation model. This makes the prior for the final density contrast highly non-Gaussian and non-linear. 

In Fourier space, the prior for the initial density contrast $\delta^\mathrm{ic}$ is a multivariate Gaussian with zero mean and diagonal covariance matrix  $\hat{S}$ corresponding to the initial cosmological power spectrum 
\begin{equation}
P(\hat{\delta}^\mathrm{ic}|\hat{S}) = \frac{1}{\sqrt{|2\pi \hat{S}|}} \exp \Bigg(-\frac{1}{2} \sum_{k,k'} \hat{\delta}^\mathrm{ic}_k \hat{S}^{-1}_{kk'}\hat{\delta}^\mathrm{ic}_{k'}\Bigg)
\label{eq:prior}
\end{equation}
where the hat denotes Fourier-space. The elements in the matrix $\hat{S}$ are fixed parameters that characterize the variance of the initial density field. This covariance matrix is diagonal in Fourier space due to the statistical homogeneity of the initial density field. The diagonal coefficients are related to the initial power spectrum $P(k)$ as
\begin{equation}
\hat{S}^{-1}_{kk} = \sqrt{\frac{P(k)}{(2\pi)^{3/2}}}.
\end{equation}
The power spectrum is chosen to follow the prescription of \citep[][]{EH98,EH99}, including BAO wiggles. 

The prior suggests physically reasonable density fields from the space of all possible states. However, the prior does not limit the space of initial conditions that is explored to match the data. If unlikely events are required to explain the data, the method explores prior regions that are unlikely.

The next step is to evolve the initial density field to the final density. We  obtain the prior for the final density contrast $\delta^\mathrm{f}$ at a scale factor $a$ by using conditional probabilities:
\begin{eqnarray}
P(\delta^\mathrm{f}) &=& \int P(\delta^\mathrm{f}, \delta^\mathrm{ic}) d\delta^\mathrm{ic} \\
&=& \int P(\delta^\mathrm{f}| \delta^\mathrm{ic})P(\delta^\mathrm{ic}) d\delta^\mathrm{ic}.
\end{eqnarray}
Since gravity is deterministic, the conditional probability is a Dirac delta of the  structure formation model $M(a,\delta^\mathrm{ic})$:
\begin{equation}
P(\delta^\mathrm{f}| \delta^\mathrm{ic}) = \prod_x \delta^D(\delta^\mathrm{f}_x - M(a,\delta^\mathrm{ic})_x).
\end{equation}

Once the gravitational model is defined, a prior distribution for the evolved density field can be obtained in two steps: first, a realization of the initial conditions is drawn from the Gaussian prior. It is then  evolved forward in time with the structure formation model. This procedure provides samples from the joint prior distribution of the initial and evolved density fields:
\begin{equation}
P(\delta^\mathrm{f},\delta^\mathrm{ic}) = P(\delta^\mathrm{ic}) \prod_x \delta^D\big(\delta^\mathrm{f}_x - M(a,\delta^\mathrm{ic})_x\big).
\end{equation}
where $x$ labels the volume elements. The evolved density field is smoothed using a Gaussian kernel with a smoothing length of $\sigma=0.5$ h$^{-1}$ Mpc.

\subsection{Likelihood}
\label{sec:likelihood}

We developed a likelihood based on the Fluctuating Gunn-Peterson Approximation \citep[FGPA][]{GPeffect}:
\begin{equation}
F_x = \exp\Bigg[-A(1+\delta_x)^\beta\Bigg]
\end{equation}
where the index $x$ labels the volume elements, $F$ is the transmitted flux fraction, $\delta$ is the density contrast and $A$ and $\beta$ are astrophysical parameters that are related to the neutral hydrogen. 

 It is common to assume that the noise in the data is Gaussian \citep{Bird11,Cisewski14,ClamatoDR1,Ozbek16,Horowitz19}. Labelling the lines of sight with an index $n$, the likelihood reads
\begin{eqnarray}
P(\delta^\mathrm{ic},\delta^\mathrm{f}|F) &=&  \prod_{n,x} \frac{1}{\sqrt{2\pi\sigma^2}} \\ \nonumber
&& \times \exp\Bigg[-\frac{\Big((F_n)_x - \exp[-A(1+\delta_x)^\beta]\Big)^2}{2\sigma^2}\Bigg]  
\label{eq:likelihood_lya}
\end{eqnarray}
where $n$ labels the lines of sight, $x$ runs over the volume elements intersected by the $n$-th line of sigth and $F$ is the observed flux. The constants $A$ and $\beta$ are astrophysical parameters that depend on the physics of the neutral hydrogen that produces the absorption lines. Although these parameters are interesting on their own because they can provide information about the intergalactic medium, they are not interesting for our cosmological analysis and we marginalize over them.

\subsection{Posterior}
\label{app:posterior_lya}
The posterior distribution of the density field can be written as
\begin{eqnarray}
P(\delta^\mathrm{f}|F) = \frac{P(F|\delta^\mathrm{f})P(\delta^\mathrm{f})}{P(F)},
\end{eqnarray}
where $P(F)$ is the normalization. In terms of the initial density field, the posterior can be written as
\begin{eqnarray}
P(\delta^\mathrm{f},\delta^\mathrm{ic}|F,\boldmath{S}) &=& \frac{P(F|\delta^\mathrm{f},\delta^\mathrm{ic})P(\delta^\mathrm{f},\delta^\mathrm{ic}|\boldmath{S})}{P(F)} \\ \nonumber
&=& P(F|\delta^\mathrm{f})\frac{P(\delta^\mathrm{ic}|\boldmath{S})}{P(F)} \prod_x \delta^D \Big[\delta^\mathrm{f}_x-M(a,\delta^\mathrm{ic})_x\Big]
\end{eqnarray}
where $x$ labels the volume elements.

By marginalizing over the final density field, we obtain the posterior that links the primordial density field with the observed flux in the \lya forest:
\begin{eqnarray}
P(\delta^\mathrm{ic}|F,\boldmath{S}) &=& P\big(F|M(a,\delta^\mathrm{ic})\big)\frac{P(\delta^\mathrm{ic}|\boldmath{S})}{P(F)} \\ \nonumber
&=& P(F|\delta^\mathrm{ic})\frac{P(\delta^\mathrm{ic}|\boldmath{S})}{P(F)} \\ 
\label{eq:posterior}
\end{eqnarray}

\subsection{Modules}
We made use of the statistical modular framework of BORG \citep{BORG}. This approach allows us to model complex data models as a Bayesian hierarchical problem. 

Since it is not efficient to sample from the high-dimensional posterior \ref{eq:posterior}, BORG makes use of the Metropolis-Hastings theorem which allows us to break the high-dimensional sampling into the sampling of the density field $\delta^\mathrm{ic}$ and the rest of the inferred quantities $\theta_i$ separately: 
\begin{eqnarray}
\Big(\delta^\mathrm{ic}\Big)^{s+1} & \curvearrowleft & P(\delta^\mathrm{ic}|\theta^s, F) \\
\theta^{s+1} & \curvearrowleft & P\Bigg((\theta^s|\Big(\delta^\mathrm{ic}\Big)^s,  F\Bigg) \\
\end{eqnarray}
where $s$ indicates the sample number, $l$ labels the volume elements and $\theta$ are the astrophysical parameters of the IGM, which are sampled using a slice sampling scheme. We obtained samples of the high-dimensional posterior distribution using a Markov Chain Monte Carlo in this block sampling scheme.

\section{Hamiltonian Monte Carlo sampling}
\label{hmc}

The inference of the density field requires inferring the amplitudes of the primordial density field at different volume elements of a regular grid, commonly between $128^3$ and $256^3$ volume elements. This implies $10^6$ to $10^7$ free parameters. Algorithms that perform a random walk, like the Metropolis-Hastings, fail in this high-dimensional parameter space. However, the Hamiltonian Monte Carlo (HMC) method provides an efficient sampling method in high-dimensional non-linear parameter spaces since it uses the information in the gradients to avoid random walks. In this section, we briefly review the HMC framework used for this work. A more detailed review, as well as its application to the large-scale structure analysis, is presented in \cite{HADES} and \cite{BORG}. 

\subsection{Algorithm}
Suppose that we want to generate samples from a probability distribution $P(\mathbf{x})$. Then the HMC scheme interprets the negative logarithm of the distribution as a potential $\psi(\mathbf{x}) = - \ln(P(\mathbf{x}))$ \citep{Duane87}.

We construct a Hamiltonian with this potential $\psi(x)$ describing the dynamics of the phase space. In order to add a kinetic term, we introduce a set of momenta $p_i$ and a mass matrix $M$ as 
\begin{eqnarray}
H(\mathbf{x},\mathbf{p}) = \sum_i\sum_j \frac{1}{2} p_i M^{-1}_{ij} p_j + \psi(x_i).
\end{eqnarray} 
The mass matrix $M$ is positive definite and affects the performance of the sampler. These masses characterize the inertia of the parameters when moving through the parameter space.

Then we can obtain the probability distribution as
\begin{eqnarray}
e^{-H} = P(\mathbf{x}) \exp\Bigg( \sum_i\sum_j \frac{1}{2} p_i M^{-1}_{ij} p_j\Bigg).
\end{eqnarray}
This ensures that the joint distribution can be split into a Gaussian distribution for the momenta $p_i$ and the distribution $P(\mathbf{x})$ to generate the samples. Therefore, the sets $\mathbf{p}$ and $\mathbf{x}$ are independent and the auxiliary momenta can be marginalized over. 

In order to generate samples from the joint distribution, we first need to draw a set of momenta from the distribution defined by the kinematic energy, which is a Gaussian distribution with covariance $\mathbf{M}$. The system evolves deterministically from an initial point in the high-dimensional parameter space, for a pseudo-time $\tau$, following the Hamilton equations: 
\begin{eqnarray}
\frac{dx_i}{dt} &=& \frac{\partial H(\mathbf{x},\mathbf{p}) }{\partial p_i} \\ 
\frac{dp_i}{dt} &=& \frac{\partial H(\mathbf{x},\mathbf{p}) }{\partial x_i} = - \frac{\partial \psi(\mathbf{x})}{\partial x_i}.
\end{eqnarray}
The new position $(\mathbf{x'},\mathbf{p'})$ in the phase space is found by integrating the Hamilton equations. Then this new point is accepted or rejected according to the original Metropolis-Hastings acceptance rule: 
\begin{eqnarray}
P_A = \min\{1, \exp[-H(\mathbf{x'},\mathbf{p'})-H(\mathbf{x},\mathbf{p})]\}. 
\end{eqnarray}
The Hamiltonian equations guarantees a unity acceptance rate since they ensure energy conservation. However, numerical inaccuracies in the integration scheme result in smaller acceptance rates. After accepting a new sample, the algorithm discards the momenta and restarts the sampling scheme by drawing a new set of momenta, which ensures the exploration of the system. Summarizing, the HMC sampling scheme is done in two steps: the first step performs a Gibbs sampling to get a set of Gaussian distributed momenta and the second step computes the deterministic dynamical trajectory on the posterior distribution.  

\section{Hamilton equations for the LSS}
The HMC exploits the gradient of the logarithmic posterior distribution to optimally explore the parameter space. Therefore, the algorithm also requires the derivatives of the posterior distribution. 

We can derive the potential in the Hamilton equations from eq. (\ref{eq:likelihood_lya}) and (\ref{eq:prior})
\begin{eqnarray}
\psi &=& -\ln\big[P\big(\delta^\mathrm{ic}|F\big) \big]\\
&=& \psi_\mathrm{likelihood}(\delta^\mathrm{ic}) +\psi_\mathrm{prior}(\delta^\mathrm{ic}),
\end{eqnarray}
where $F$ is the data and
\begin{eqnarray}
\psi_\mathrm{prior}(\delta^\mathrm{ic}) &=& \frac{1}{2} \sum_{xy} \delta^\mathrm{ic}_x S^{-1}_{xy}  \delta^\mathrm{ic}_y, \\ 
\psi_\mathrm{likelihood}(\delta^\mathrm{ic}) &=& \sum_n \sum_x^N \frac{\Big((F_n)_x -  \exp[A(1+M(a,\delta^\mathrm{ic}))_x^\beta]\Big)^2}{2\sigma^2} \nonumber  \\ 
&& + \frac{1}{2^N} \sum_n \ln(2\pi\sigma^2)^N \label{eq:likelihood}
\end{eqnarray}
where $n$ runs over the lines of sight, $x$ and $y$ label the volume elements and $N$ is the number of voxels of the $n$-th line of sight. 

The forces are computed by differentiating with respect to $\delta^\mathrm{ic}$
\begin{eqnarray}
\frac{\partial \psi(\delta^\mathrm{ic})}{\partial \delta^\mathrm{ic}_x} = \frac{\partial \psi_\mathrm{prior}(\delta^\mathrm{ic})}{\partial \delta^\mathrm{ic}_x} + \frac{\partial \psi_\mathrm{likelihood}(\delta^\mathrm{ic})}{\partial \delta_x^\mathrm{ic}}.
\end{eqnarray}
The prior gradient is
\begin{eqnarray}
\frac{\partial \psi_\mathrm{prior}(\delta^\mathrm{ic})}{\partial \delta^\mathrm{ic}_x} = \sum_y S^{-1}_{xy} \delta^\mathrm{ic}_y
\end{eqnarray}
and the corresponding likelihood term 
\begin{eqnarray}
\frac{\partial \psi_\mathrm{likelihood}(\delta^\mathrm{ic}))}{\partial \delta^\mathrm{ic}_x} &=&  \sum_n \frac{(F_n)_x - \exp[-A(1+M(a,\delta^\mathrm{ic})_x)^{\beta}]}{\sigma^2} \nonumber \\ 
&\times& A \beta \big(1+M(a,\delta^\mathrm{ic})_x\big)^{\beta-1} \\ \nonumber
&\times& \exp\Big[-A\big(1+M(a,\delta^\mathrm{ic})_x\big)^{\beta}\Big] \\ \nonumber 
&\times&\frac{\partial M(a,\delta^\mathrm{ic})_x }{\partial \delta^\mathrm{ic}_x} .
\end{eqnarray}
The parameters of the likelihood $A$, $\beta$ and the noise variance $\sigma$ are sampled using a slice sampler. The equations of motion are integrated using a leapfrog integration scheme \cite{Duane87}, which is exactly reversible and symplectic, maintaining the detailed balance of the Markov chain.

\section{Pearson's correlation coefficient}
\label{app:pearson}
The Pearson's correlation coefficient is a measurement of the linear correlation between two variables. It has a value between 1 and -1, with 1 indicating linear positive correlation and -1 anticorrelation. The Pearson's coefficient is defined as
\begin{equation}
r_{xy} = \frac{\sum_i (x_i-\langle x\rangle)(y_i-\langle y\rangle)}{\Big[\sum_i (x_i - \langle x\rangle)^2\Big]^{1/2}\Big[\sum_i (y_i - \langle y\rangle)^2\Big]^{1/2}}
\end{equation}
where $i$ labels the samples and $\langle x\rangle$ indicates the sample mean of $x$.

\end{document}